\documentclass[%
 reprint,
 amsmath,amssymb,
 aps,
pra,
showkeys
]{revtex4-2}

\usepackage{graphicx}
\usepackage{dcolumn}
\usepackage{bm}
\usepackage{braket}
\usepackage{xcolor}
\usepackage{colortbl}

\begin{document}

\preprint{APS/123-QED}

\title{Frustration-Enhanced Quantum Annealing Correction Models with Additional Inter-replica Interactions}

\author{Tomohiro Hattori$^1$}
\affiliation{$^1$ Graduate School of Science and Technology, Keio University, 3-14-1 Hiyoshi, Kohoku-ku, Yokohama-shi, Kanagawa 223-8522, Japan}
\author{Shu Tanaka$^{1,2,3,4}$\thanks{shu.tanaka@appi.keio.ac.jp}}
\affiliation{$^2$ Department of Applied Physics and Physico-Informatics, Keio University, 3-14-1 Hiyoshi, Kohoku-ku, Yokohama-shi, Kanagawa 223-8522, Japan \\
$^3$ Keio University Sustainable Quantum Artificial Intelligence Center (KSQAIC), Keio University, Tokyo 108-8345, Japan \\
$^4$ Human Biology-Microbiome-Quantum Research Center (WPI-Bio2Q), Keio University, Tokyo 108-8345, Japan
}

\date{\today}
\begin{abstract}
Quantum annealing correction (QAC) models provide a promising approach for mitigating errors in quantum annealers.
Previous studies have established that QAC models are crucial for ensuring the robustness of the ground state of the Ising model on hardware.
In this work, the effects of QAC models incorporating replicas with additional interactions, specifically, the {\it penalty spin model} and the {\it stacked model}, are investigated for problems characterized by a small energy gap between the ground and first excited states during quantum annealing, a well-known bottleneck to reaching the ground state.
The results demonstrate that these QAC models can obtain the optimal solution within short annealing times by exploiting diabatic transitions, even for problems with a small energy gap.
These findings highlight the potential of QAC models as practical near-term algorithms for hardware subject to runtime limitations and control noise.
\end{abstract}

\maketitle

\section{\label{sec:introduction}Introduction}
Quantum Annealing (QA)~\cite{kadowaki1998quantum, farhi2000quantum, mcgeoch2014adiabatic, tanaka2017quantum, albash2018adiabatic, chakrabarti2023quantum} has the potential to solve combinatorial optimization problems with both high accuracy and efficiency. 
Recently, various QA-based algorithms have been developed for practical use, including size reduction methods~\cite{Karmi2017, Karimi2017_2, Irie2021Hybrid, Atobe_2022, kikuchi2023hybrid, hattori2025advantages, hattori2025impact}, quantum-classical hybrid methods~\cite{hirama2023efficient, kanai2024annealing, hattori2026improving}, and error mitigation methods~\cite{raymond2025quantum, shingu2024quantummitigation}.
Moreover, QA and QA-inspired methods have been applied to diverse domains, including advertisement optimization~\cite{tanahashi2019application}, black-box optimization~\cite{FMA2020, FMA2022,photonic_laser2022, matsumori2022application, FMA2023, couzinie2025machine, tamura2026black}, computer-aided engineering~\cite{harris2018phase, king2018observation, endo2022phase, honda2024development, xu2025quantum}, and finance~\cite{rosenberg2015solving_trading, grant2021benchmarking_portofolio}.

The adiabatic theorem~\cite{kato1950adiabatic} guarantees that the ground state of the Ising model, which corresponds to the optimal solution of the combinatorial optimization problems~\cite{lucas2014ising}, is obtained at the final time of QA with a sufficiently long annealing time.
Previous studies~\cite{kato1950adiabatic, mcgeoch2014adiabatic, tanaka2017quantum, albash2018adiabatic, chakrabarti2023quantum} have discussed the adiabatic condition, which determines the required annealing time to obtain the ground state of the Ising model.
The adiabatic condition guarantees instantaneously following the ground state of the system during QA with an annealing time that scales with the inverse square of the minimum energy gap between the ground state and the first excited state.
However, bottlenecks such as a small energy gap~\cite{altshuler2010anderson}, hardware control noise~\cite{pearson2019analog}, and decoherence~\cite{albash2015decoherence} hinder the achievement of high-quality solutions.
Therefore, algorithms are needed to overcome these bottlenecks.

Expanding the energy gap is one promising approach.
Previous studies have demonstrated improvements in the scaling of the energy gap, both analytically and numerically~\cite{altshuler2010anderson, seki2012quantum, somma2012quantum, susa2018exponential, susa2018quantum,adame2020inhomogeneous, albash2021diagonal, Zaborniak:2021aql, feinstein2024effects, ghosh2026enhancement}.
Although numerous methods have achieved exponential improvements in energy-gap scaling, significant difficulties remain in their practical implementation on quantum hardware and in their limited applicability under specific enhancement conditions.

Exploiting diabatic transitions during QA is another promising approach for overcoming the small energy gap~\cite{crosson2021prospects, cote2023diabatic, feinstein2025robustness, hattori2026improving}.
Diabatic QA, which allows diabatic transitions during QA, has been shown to benefit both combinatorial optimization and sampling tasks~\cite{crosson2021prospects}.
Moreover, annealing schedule optimization is a promising way to exploit diabatic transitions during QA~\cite{cote2023diabatic, hattori2026improving}.
Although the transferability of annealing schedules reduces the cost of the schedule optimization~\cite{hattori2026improving}, the applicability remains limited for similar problem instances.
These methods exploiting diabatic transitions are important as near-term algorithms, since diabatic transitions are inevitable in current hardware.

The parallel QA has been proposed to effectively use the idle quantum bits (qubits), thereby improving the time-to-solution (TTS) in real quantum annealers~\cite{Pelofske_2022}.
Also, previous studies have proposed the quantum annealing correction (QAC) models to enhance the robustness of the ground state of the Ising model against control errors~\cite{pudenz2014error, pudenz2015quantumerror, vinci2015quantumerror, vinci2016nested, matsuura2016mean, mishra2016performance, matsuura2017quantum, vinci2018scalable, matsuura2019nested, pearson2019analog, bennett2023using, hino2024physical}, and demonstrated better performance than parallel QA, which is equivalent to the classical model in these previous studies.
In the previous studies, errors in a random error Ising model were suppressed~\cite{pudenz2014error} by introducing additional interactions between each spin and a {\it penalty spin}.
A penalty spin acts as a slack spin that imposes an energetic penalty when spins across replicas are not aligned, where a replica is a spatially parallelized copy of the Ising model. 
This QAC model is referred to as the {\it penalty spin model}.
Consequently, the additional interactions cause the spins in each replica to align in the same direction, thereby protecting the quantum state from errors.
The penalty spin model is particularly effective when the success probability of each replica is relatively high, as it resembles the majority-vote scheme in coding theory.
Analytical studies of penalty spin models have also been actively pursued~\cite{matsuura2016mean, matsuura2017quantum}.

The nested quantum annealing correction (NQAC) model~\cite{vinci2016nested, matsuura2019nested} is an extension of QAC models that introduces a nested structure to enhance both error resilience and solution quality by adding ferromagnetic interactions that align the replicas in the same direction.
This structure mitigates spin flips, chain breaks, and other hardware-induced errors, resulting in stable success probabilities that are largely independent of the number of replicas.
The NQAC model has been shown to be effective even for Boltzmann sampling via QA~\cite{li2019improved, li2020limitations}.
The effectiveness of antiferromagnetic interactions has been validated in the NQAC model.
However, the mechanism underlying this performance improvement remains unclear.

Another promising QAC approach is the stacked model, which utilizes replicas of the Ising model on hardware and incorporates additional interactions between replicas~\cite{young2013adiabatic, bennett2023using, hino2024physical}. 
Previous studies demonstrated the robustness of the ground state of the Ising model for both the Sherrington--Kirkpatrick model and the antiferromagnetic chain~\cite{bennett2023using, hino2024physical}. 
Additionally, they investigated in detail how and when antiferromagnetic interactions between replicas can effectively reduce bit-flip errors, and they estimated the extent of error reduction achieved.
However, the dynamics of QA and the performance on actual quantum annealers have not been studied in instances with a small energy gap between the ground state and the first excited state.
Adding slack spins may hinder the solution quality by increasing the number of spins, as suggested in the previous studies~\cite{kikuchi2023hybrid, hattori2025advantages}.
Therefore, it is crucial to investigate the properties of the QAC models and whether they hinder the performance.

In this study, we examined QAC models applied to problems characterized by an exponentially decreasing minimum energy gap.
 Previous studies have discussed the effects of adding ferromagnetic interactions in many QAC models. 
Such approaches can be interpreted as error-correction schemes based on majority voting. 
In contrast, relatively few studies have investigated QAC models that incorporate antiferromagnetic interactions.
While the robustness of the ground state in formulations with added antiferromagnetic interactions has been examined, experimental demonstrations of success probability on actual QA hardware remain lacking, and it is still unclear for which classes of problems such models are most effective.

In this context, the present work aims to explore practical ways to utilize QAC models by leveraging their intrinsic decoding properties.
In QAC models, the optimal solution may appear in certain excited eigenstates after decoding, thus, even if the quantum state does not reach the ground state, the optimal solution can still be obtained.
Therefore, under conditions where the state is easily excited due to the small energy gap, decoherence, or control noise, QAC models can still yield the optimal solution.
To evaluate the effects of the QAC models, we used the frustrated ring model as a benchmark problem, since it exhibits spin-glass bottlenecks where the minimum energy gap between the ground state and the first excited state during QA decreases exponentially with system size~\cite{roberts2020noise}.

The remainder of this paper is organized as follows.
Section~\ref{sec: method} introduces QA and QAC models.
Section~\ref{sec: settings} describes the frustrated ring model, experimental setup, numerical methods, and evaluation metrics.
Section~\ref{sec: result} analyzes the effects of the QAC models in experiments and numerical simulations, focusing on success probability and the properties of the energy spectrum during QA from the perspective of diabatic QA.
Finally, Section~\ref{sec: conclusion} presents our conclusions and outlines future directions. 

\section{\label{sec: method}Method}
\begin{figure*}
    \centering
    \includegraphics[scale=0.35,clip]{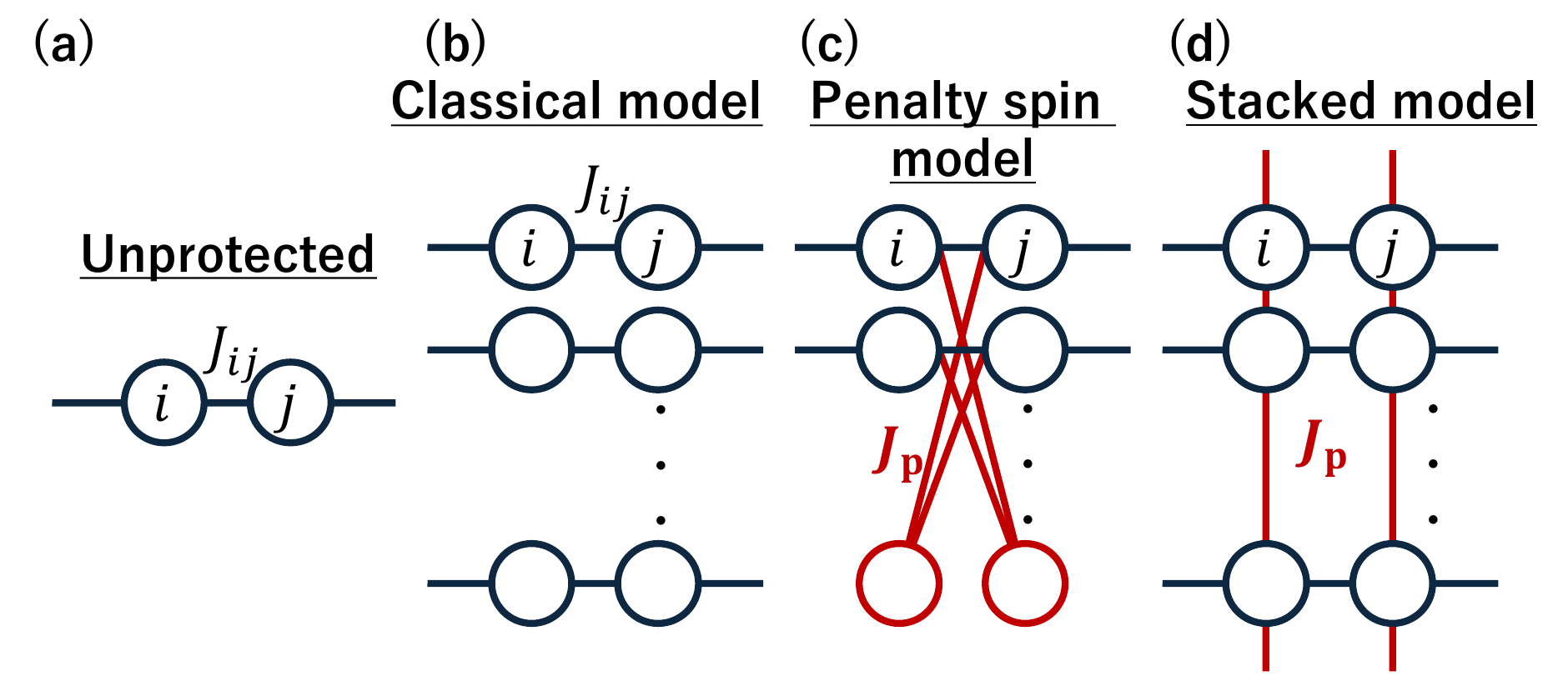}
    \caption{Schematic diagrams of the QAC models:
    (a) the unprotected model,
    (b) the classical model,
    (c) the penalty spin model,
    (d) the stacked model.
     The unprotected model is equivalent to the Ising model with conventional QA, and the classical model uses replicas without interactions between replicas, which is equivalent to the parallel QA.
    The penalty spin model and the stacked model use replicas with additional interactions.
    The circles represent the spins, and the solid lines represent the interactions.
    The red circles represent the slack spins, and the red solid lines represent the additional interactions.
    }
    \label{fig: QAC_method}
\end{figure*}

In QA, the optimal solution of a combinatorial optimization problem corresponds to the ground state of the Ising model. 
The Hamiltonian of the QA is expressed as follows:
\begin{align}
    \label{eq: H_QA}
    \mathcal{H}(t)=A(t)\mathcal{H}_\mathrm{q}+B(t)\mathcal{H}_\mathrm{p},\quad 0\leq t\leq\tau,
\end{align}
where $\tau$ is the annealing time, $\mathcal{H}_\mathrm{q}$ is the driver Hamiltonian, $\mathcal{H}_\mathrm{p}$ is the problem Hamiltonian, and $A(t)$ and $B(t)$ are real-valued functions called annealing schedules.
The annealing schedules satisfy $A(0)\gg B(0)$ and $A(\tau)\ll B(\tau)$.
Many combinatorial optimization problems can be formulated as an Ising model by mapping their optimal solutions to the combinatorial optimization problems and the ground state of the Ising model~\cite{lucas2014ising}.
In QA, the problem Hamiltonian is represented by the Ising model.
A transverse field is generally used as the driver Hamiltonian.
QA ideally aims for the quantum state to follow the instantaneous ground state of the system.
Therefore, the initial state is the ground state of $\mathcal{H}_\mathrm{q}$.

The transverse-field and problem Hamiltonians are respectively expressed as follows:
\begin{align}
    \label{eq: tfHamiltonian}
    \mathcal{H}_\mathrm{q}&=-\sum_{i=1}^N\sigma_i^x,\\
    \label{eq: pHamiltonian}
    \mathcal{H}_\mathrm{p}&= -\sum_{i=1}^Nh_i\sigma_i^z-\sum_{1\leq i<j\leq N}J_{i,j}\sigma_i^z\sigma_j^z,
\end{align}
where $\sigma_i^\alpha$ ($\alpha =x, y, z$) is the $\alpha$-component of the Pauli matrix on vertex $i$ and $N$ is the number of spins.
Here, $h_i$ and $J_{i,j}$ are the longitudinal magnetic fields at spin $i$ and the interactions between spin $i$ and $j$, respectively.
The initial state is the equal superposition state of all computational basis states, which corresponds to the ground state of $\mathcal{H}_\mathrm{q}$.
The formulation of $\mathcal{H}_\mathrm{p}$ in Eq.~\eqref{eq: pHamiltonian} is the standard form used in QA, which we call the {\it unprotected model} as shown in Fig.~\ref{fig: QAC_method}~(a).

The required number of spins is $N\times K$ in the QAC models.
The unpenalized parallel Ising model, referred to as the {\it classical model}~\cite{pudenz2014error, pudenz2015quantumerror}, was also proposed as a form of parallel QA in a previous study~\cite{Pelofske_2022}.
A schematic diagram of the classical model is shown in Fig.~\ref{fig: QAC_method}~(b).
We define the Hamiltonian of the $k$th replica of the QAC model as
\begin{align}
    \label{eq: kth_Hp}
    \mathcal{H}_\mathrm{p}^{(k)} =& -\sum_{i=1}^Nh_i\sigma_{i,k}^z-\sum_{1\leq i<j\leq N}J_{i,j}\sigma_{i,k}^z\sigma_{j,k}^z.
\end{align}
Here, $\sigma_{i,k}^\alpha$ ($\alpha =x, y, z$) denotes the $\alpha$-component of the Pauli matrix on vertex $i$ in the $k$th replica.
In the classical model, the Hamiltonian is given by 
\begin{align}
    \label{eq: Cmodel}
    \mathcal{H}_\mathrm{C}=\sum_{k=1}^{K}\mathcal{H}_\mathrm{p}^{(k)},
\end{align}
where $K$ denotes the number of replicas.

The {\it penalty spin model} has been proposed as a QAC model that employs replicas and additional interactions~\cite{pudenz2014error, pudenz2015quantumerror, vinci2015quantumerror, vinci2016nested, pearson2019analog}.
A schematic diagram of the penalty spin model is shown in Fig.~\ref{fig: QAC_method}~(c).
The penalty spin model protects the quantum states from errors such as decoherence and bond chaos.
In the penalty spin model, the additional interactions between each spin and the {\it penalty spin} are introduced.
A penalty spin acts as a slack spin that imposes an energetic penalty when the spins across replicas are not aligned. 
The formulation of the penalty spin model is as follows:
\begin{align}
    \label{eq: PSmodel}
    \mathcal{H}_\mathrm{ps}=\sum_{k=1}^{K-1}\mathcal{H}_\mathrm{p}^{(k)} - J_\mathrm{p}\sum_{i=1}^N\sum_{k=1}^{K-1}\sigma_{i,k}^z\sigma_{i,K}^z,
\end{align}
where $\mathcal{H}_\mathrm{p}^{(k)}$ is the same as Eq.~\eqref{eq: kth_Hp}, and $J_\mathrm{p}$ is the interaction strength between each spin and the penalty spin.
As shown in Eq.~\eqref{eq: PSmodel}, the penalty spin model requires $N\times K$ spins, consisting of $K-1$ replicas together with $N$ penalty spins.
We decode the final state by selecting, among the $K-1$ problem replicas and penalty spins, the replica with the lowest energy in the penalty-spin model.
In the penalty spin model, spins at the same site align regardless of the sign of $J_\mathrm{p}$, while the alignment of the penalty spins depends on the sign of $J_\mathrm{p}$.

In this study, we consider the stacked model as a QAC model, which spatially parallelizes replicas of $\mathcal{H}_\mathrm{p}$ with additional inter-replica interactions, as proposed in the previous study~\cite{young2013adiabatic, bennett2023using, hino2024physical}.
A schematic diagram of the stacked model is shown in Fig.~\ref{fig: QAC_method}~(d).
We investigate two boundary conditions in this model: open and periodic.
The Hamiltonian of the stacked model with open boundary conditions is given by 
\begin{align}
    \label{eq: stacked_obc}
    \mathcal{H}_\mathrm{stacked}^\mathrm{(obc)} =\sum_{k=1}^{K}\mathcal{H}_\mathrm{p}^{(k)}-J_\mathrm{p}\sum_{i=1}^N\sum_{k=1}^{K-1}\sigma_{i,k}^z\sigma_{i,k+1}^z,
\end{align}
where $\mathcal{H}_\mathrm{p}^{(k)}$ is the same as Eq.~\eqref{eq: kth_Hp}, $J_\mathrm{p}$ is the interaction strength between replicas.
The Hamiltonian of the stacked model with periodic boundary conditions is given by
\begin{align}
    \label{eq: stacked_pbc}
    \mathcal{H}_\mathrm{stacked}^\mathrm{(pbc)} =\sum_{k=1}^{K}\mathcal{H}_\mathrm{p}^{(k)}-J_\mathrm{p}\sum_{i=1}^N\sum_{k=1}^K\sigma_{i,k}^z\sigma_{i,k+1}^z.
\end{align}
Here, $\sigma_{i,K+1}=\sigma_{i,1}$ for arbitrary $i$ due to the periodic boundary condition.
For $J_\mathrm{p}>0$ (ferromagnetic interactions), the additional interactions align spins among different replicas in the same direction.
In contrast, the additional interactions with $J_\mathrm{p}<0$ (antiferromagnetic interactions) align spins of nearest-neighbor replicas in opposite directions.

When we apply the QAC models, we replace $\mathcal{H}_\mathrm{p}$ in Eq.~\eqref{eq: H_QA} with $\mathcal{H}_\mathrm{ps}$ shown in Eq.~\eqref{eq: PSmodel} or $\mathcal{H}_\mathrm{stacked}^\mathrm{(obc)}{}$,$\mathcal{H}_\mathrm{stacked}^\mathrm{(pbc)}$ shown in Eqs.~\eqref{eq: stacked_obc}, \eqref{eq: stacked_pbc}.
In these QAC models using $K$ replicas, the transverse fields are expressed as follows:
\begin{align}
    \label{eq: tf_QAC}
    \mathcal{H}_\mathrm{q}^{(K)}=-\sum_{k=1}^K\sum_{i=1}^{N}\sigma_{i,k}^x.
\end{align}
 Here, the initial state is the ground state of the transverse fields defined in Eq.~\eqref{eq: tf_QAC}, which is easy to prepare on hardware.

Decoding strategies for QAC models have been discussed in previous studies~\cite{pudenz2014error, pudenz2015quantumerror, vinci2015quantumerror}.
These previous studies demonstrated that the energy-minimization strategy is more effective than the majority-vote strategy, which determines spin configurations by a majority rule across parallelized replicas.
Therefore, we adopt the energy-minimization strategy for decoding QAC models.
In this strategy, the energy of each replica is calculated, and the replica with the lowest energy is selected.

\section{\label{sec: settings}Settings}
\begin{figure}[t]
    \centering
    \includegraphics[clip,scale=0.35]{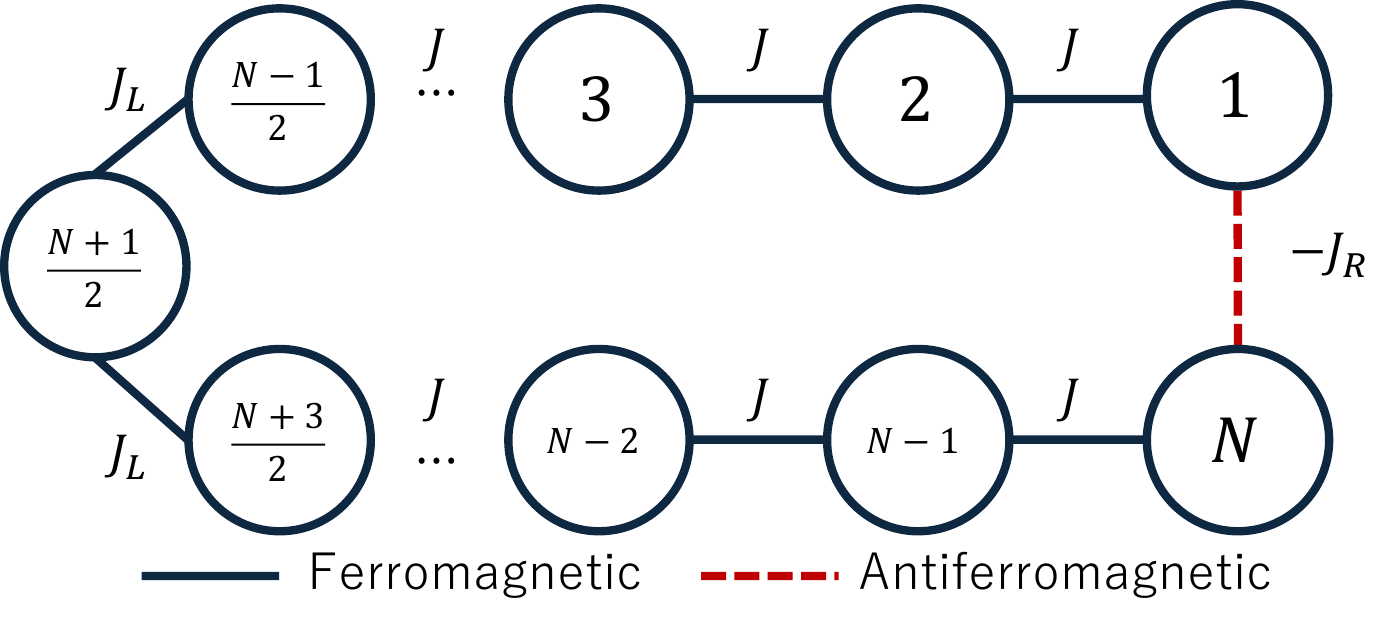}
    \caption{
    Schematic diagram of the frustrated ring model with an odd number of spins $N$.
    The circles represent the spins, the solid lines represent the ferromagnetic interactions, and the red dashed line represents the antiferromagnetic interaction.
    The interaction strengths are defined in Eq.~\eqref{eq: interaction_fr}.
    }
    \label{fig: FrustratedRingModel}
\end{figure}
\begin{figure}[t]
    \centering
    \includegraphics[scale=1,clip]{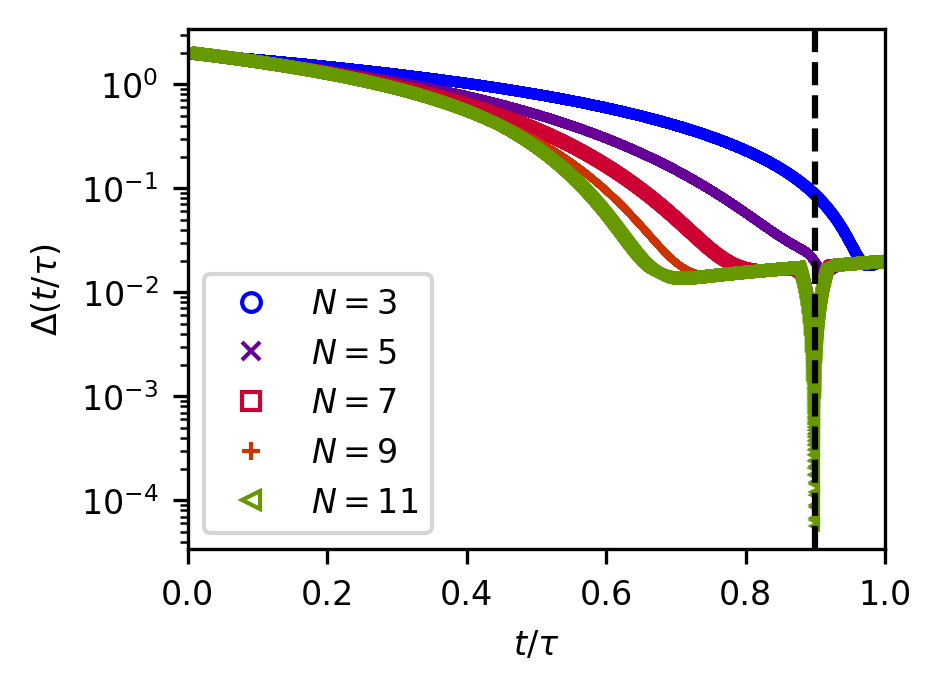}
    \caption{
    Energy gap $\Delta(t/\tau)$ between the ground state and the first excited state during QA for the frustrated ring model after removing the degeneracies by adding longitudinal magnetic fields $h_1=0.01$, shown in Eq.~\eqref{eq: frustrated_ring_model_without_degenerate}.
    The interactions of the frustrated ring model are $J=1$, $J_\mathrm{L}=0.5$ and $J_\mathrm{R}=0.45$.
    The black dashed line indicates the time at which the energy gap closes at $N\rightarrow\infty$, obtained analytically from the frustrated ring model with $h_1=0$ in Eq.~\eqref{eq: frustrated_ring_model_without_degenerate}.
    }
    \label{fig: FRgap}
\end{figure}

In this study, we adopted the frustrated ring model as a benchmark problem because its sparsity results in minimal embedding overhead.
Figure~\ref{fig: FrustratedRingModel} shows a schematic diagram of the frustrated ring model with an odd number of spins $N$.
The frustrated ring model is a benchmark problem known to exhibit a spin-glass bottleneck during the QA~\cite{wang2025exponential}.

The spin-glass bottleneck, a well-known limitation in QA, prevents the quantum state from remaining the instantaneous ground state of the total Hamiltonian.
Therefore, obtaining the ground state of the frustrated ring model with conventional QA is difficult.

In this study, to eliminate trivial degeneracies for simplicity, we applied a small longitudinal magnetic field at site $1$, setting $h_1=0.01$.
The frustrated ring model used in this study is therefore expressed as follows:
\begin{align}
    \label{eq: frustrated_ring_model_without_degenerate}
    \mathcal{H}_\mathrm{fr}=-\sum_{i=1}^N J_i\sigma_i^z\sigma_{i+1}^z - h_1\sigma_1^z.
\end{align}
The interactions are defined as follows:
\begin{align}
    \label{eq: interaction_fr}
    J_i =
    \begin{cases}
    -J_\mathrm{R} & \text{if } i = N \\
    J_\mathrm{L} & \text{if } i = \frac{1}{2}(N \pm 1) \\
    J   & \text{otherwise}
\end{cases},
\end{align}
where the interactions satisfy $0 < J_\mathrm{R} < J_\mathrm{L} < J = 1$, $N$ is odd, and the periodic boundary condition is imposed, such that $i=N+1$ corresponds to $i=1$.
Interactions between nearest-neighbor spins are ferromagnetic, and the interaction between spin $1$ and spin $N$ is antiferromagnetic, which introduces frustration in the system.

The frustrated ring model dealt with in the previous study~\cite {roberts2020noise} has no longitudinal magnetic fields, which is $h_1=0$ in Eq.~\eqref{eq: frustrated_ring_model_without_degenerate}.
Following Ref.~\cite{roberts2020noise}, we set the parameters of the frustrated ring model in the spin-glass bottleneck region to $J_\mathrm{R}=0.45$ and $J_\mathrm{L}=0.5$.
Figure~\ref{fig: FRgap} shows the energy gap between the ground state and the first excited state during QA for the frustrated ring model shown in Eq.~\eqref{eq: frustrated_ring_model_without_degenerate}.
As seen in Fig.~\ref{fig: FRgap}, the small-gap properties persist even after removing the degeneracies.
Therefore, the frustrated ring model remains difficult to solve using both classical metaheuristics and QA.

In this setting, the ground state of the frustrated ring model is the all-up state, and the first excited state of the frustrated ring model is the all-down state, whereas these two states are degenerate without the longitudinal magnetic field.
Under this setting, classical metaheuristics using a single-spin flip, such as simulated annealing~\cite{kirkpatrick1983optimization} and gradient descent methods, are easily trapped in local minima due to the large Hamming distance between the ground state and the first excited state.

We evaluated the performance of the QAC models on a quantum annealer.
We used \verb|Advantage2_system1.10| as the quantum annealer~\cite{D-Wave_params}. 
The quantum annealer consists of $4594$ qubits arranged on the Zephyr graph~\cite{D-Wave_zephr}.
Since the Zephyr graph is a two-dimensional lattice with open boundary conditions, the stacked model, whose unprotected model is a one-dimensional chain with open boundary conditions, is embeddable without embedding overhead.
The embedding overhead is defined as the number of spins in one chain.
The total number of spins of the frustrated ring model on the Zephyr graph is $O(N)$, which is significantly smaller than the total number of spins of complete graphs $O(N^2)$, indicating that the frustrated ring model has a smaller average chain length.

\begin{figure}[t]
    \centering
    \includegraphics[scale=1.0,clip]{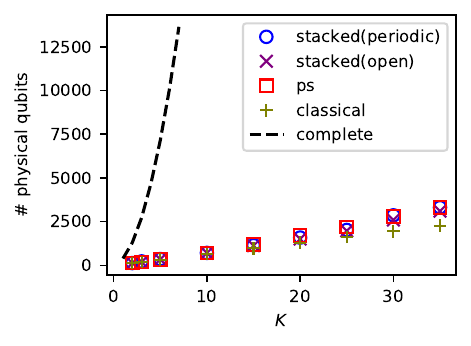}
    \caption{Number of physical spins on the Zephyr graph when embedding the frustrated ring model with QAC models.
    The black dashed line shows the number of physical qubits by using clique embedding.}
    \label{fig: number_of_physical_spin}
\end{figure}
 Figure~\ref{fig: number_of_physical_spin} shows the number of physical spins of QAC models on the Zephyr graph.
Here, the lowest number of physical qubits is used by employing minor embedding~\cite{Choi_2008, Choi_2011} among $1000$ runs.
Even on the Zephyr graph of the QAC models, the embedding overhead is much smaller than the complete graph.
The effects of embedding on each QAC model are in Appendix~\ref{appendix: influence_of_chain}.

We set the number of spins in the unprotected model $N=11, 21, 31, 41, 51, 61$ and the number of replicas to $K=1, 2, 3, 5, 10, 15, 20, 25, 30, 35$, considering the overhead arising from differences in the boundary conditions between the Zephyr graph and the QAC models.
The annealing time was set to $\tau = 1, 20, 100, 1000, 2000~\mathrm{\mu s}$.
In the analysis of the dependence of the other parameters, the annealing time was set to $\tau = 20~\mathrm{\mu s}$, which is the default annealing time.
We examined the dependence of performance on annealing time, additional interactions in the QAC models, and the number of spins.
The quantum annealer was executed with \verb|num_reads| $= 100$ and minor embedding was employed~\cite{Choi_2008, Choi_2011}.
All other parameters were set to their default values.

Next, we numerically examined the effectiveness of the QAC models for the frustrated ring model by solving the time-dependent Schr\"{o}dinger equation using QuTiP~\cite{QuTiP1, QuTiP2}.
The annealing schedules are set to $A(t)=1-t/\tau$ and $B(t)=t/\tau$ for simplicity.
We set the number of spins to $N=3$ to investigate the dependence of the number of replicas in the numerical analysis.
The Schr\"{o}dinger equation is written as follows:
\begin{align}
    \label{eq: SE}
    \mathrm{i}\frac{\partial}{\partial t}\ket{\psi(t)}=\mathcal{H}(t)\ket{\psi(t)}.
\end{align}
Here, Planck units are used.

We evaluated the performance of the QAC models by using the success probability.
The success probability is expressed as follows:
\begin{align}
    \label{eq: instantaneous_probability}
    p(\tau)=\bra{\psi(\tau)}P_\mathrm{s}\ket{\psi(\tau)},
\end{align}
where $\ket{\psi(\tau)}$ represents the quantum state at a time $\tau$.
The failure probability is defined as the complement of the success probability.
The projection operator $P_\mathrm{s}$ is defined as
\begin{align}
    \label{eq: phis}
    P_\mathrm{s}=\sum_{i\in S_\mathrm{s}}\ket{\phi_i}\bra{\phi_i},
\end{align}
where $\ket{\phi_i}$ is the $i$th eigenstate of the Hamiltonian of the QAC models and $S_\mathrm{s}$ is the set of indices of states including the optimal solution after decoding in the QAC model.

We examined the energy spectrum through numerical exact diagonalization.
To reveal the properties of the QAC models, we compared the unprotected model, the classical model, and the QAC models with different numbers of replicas and interaction strengths by using these metrics.

\section{\label{sec: result}Results}
We evaluated the performance of QAC models, including the stacked models with both periodic and open boundary conditions, as well as the penalty spin model, using an actual quantum annealer and numerical simulation.
Section~\ref{subsec: dwave_experiment} presents the performance of QAC models on the actual quantum annealer, and Section~\ref{subsec: numerical_analysis} shows the numerical results.

\subsection{Experiment on an actual quantum annealer\label{subsec: dwave_experiment}}
\begin{figure*}[t]
    \centering
    \includegraphics[clip,scale=1.0]{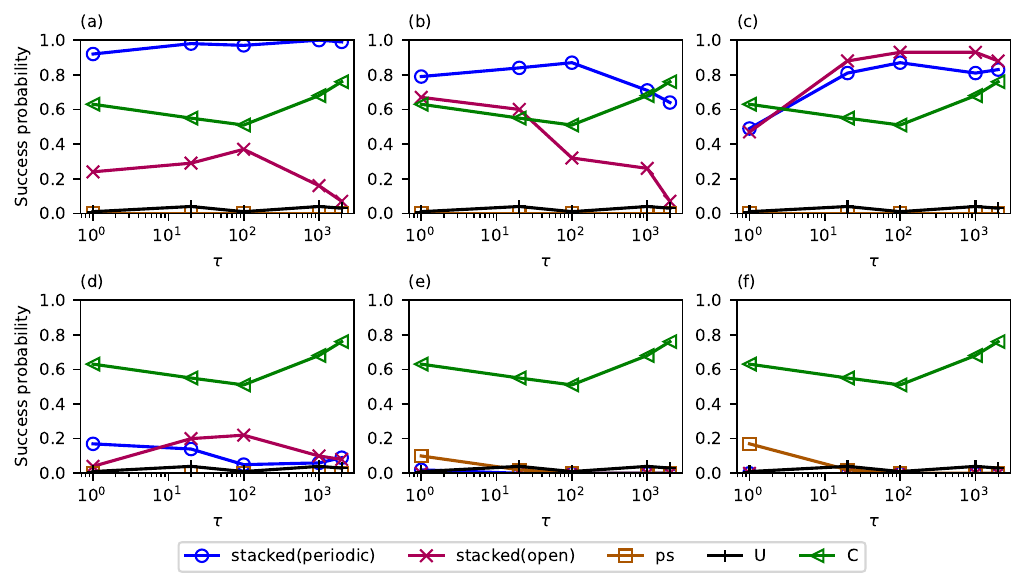}
    \caption{Final success probability as a function of annealing time $\tau$.
    The horizontal axis is shown on a logarithmic scale.
    The problem size is $N = 61$.
    The number of replicas is $K=15$.
    The black and green lines represent the success probabilities of the unprotected and classical models, respectively.
    Each panel shows the results with the different interaction strength $J_\mathrm{p}$: (a) $J_\mathrm{p}=-1$, (b) $J_\mathrm{p}=-0.1$, (c) $J_\mathrm{p}=-0.01$, (d) $J_\mathrm{p}=0.01$, (e) $J_\mathrm{p}=0.1$ and (f) $J_\mathrm{p}=1$.
    Solid lines between points are provided as a guide to the eye.}
    \label{fig: dwave: Jp_dependence_success_probability}
\end{figure*}
\begin{figure}[t]
    \centering
    \includegraphics[clip,scale=1.0]{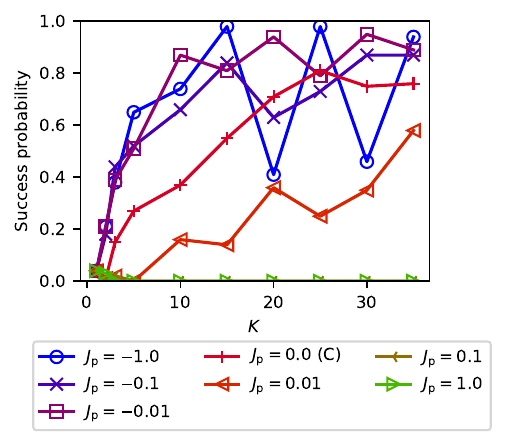}
    \caption{The final success probability as a function of the number of replicas $K$ with the periodic-boundary stacked model.
    The problem size is $N = 61$.
    The annealing time is $\tau=20~\mathrm{\mu s}$.
    The results at $J_\mathrm{p}=0$ correspond to the classical model.
    The dashed black line represents the success probability of the unprotected model.
    Solid lines between points are provided as a guide to the eye.
    }
    \label{fig: dwave: success_probability_K}
\end{figure}
\begin{figure}[t]
    \centering
    \includegraphics[clip,scale=1.0]{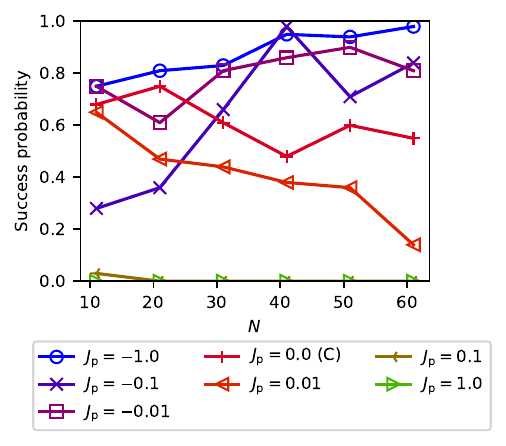}
    \caption{Scaling of the final success probability of the periodic-boundary stacked model.
    The number of replicas is fixed to $K=15$.
    The annealing time is $\tau=~20\mathrm{\mu s}$.
    The results at $J_\mathrm{p}=0$ correspond to the classical model.
    The dashed black line represents the success probability of the unprotected model.
    Solid lines between points are provided as a guide to the eye.
    }
    \label{fig: dwave: success_probability_scaling}
\end{figure}
We first examined the effect of additional interactions on the success probability.
Figure~\ref{fig: dwave: Jp_dependence_success_probability} shows the final success probability as a function of the annealing time $\tau$ for the frustrated ring model after removing degeneracies with $N=61$ and $K=15$.
The number of replicas is selected as the efficient number of qubits by time-to-solution (TTS) considering the efficiency of the qubit resource in Appendix~\ref{appendix: TTS}.

From the perspective of efficient resource utilization, it is also important to consider parallelizing the QAC model itself.
We therefore included in Appendix~\ref{appendix:parallel_QAC} a comparison between a parallelized implementation of the QAC model and a single QAC model that uses the same number of logical qubits.

Figure~\ref{fig: dwave: Jp_dependence_success_probability} demonstrates that the QAC model enhances performance compared to the unprotected model, whereas ferromagnetic interactions degrade it.
Adding interactions increases the energy scale of the system, which can hinder exploration under a fixed transverse-field scale. 
However, Fig.~\ref{fig: dwave: Jp_dependence_success_probability} shows that the additional antiferromagnetic interactions broaden the accessible low-energy states, facilitating exploration, enabling high-quality solutions to be obtained with shorter annealing times.
We next examine each model in detail.

In the periodic-boundary stacked model shown in Figs.~\ref{fig: dwave: Jp_dependence_success_probability}~(a)-(c), introducing antiferromagnetic interactions leads to a higher success probability than that of the classical model.
This enhancement is observed regardless of the magnitude of the antiferromagnetic interactions within the $J_\mathrm{p}$ range examined, $J_\mathrm{p}=-1, -0.1, -0.01$.
By contrast, ferromagnetic interactions reduce the success probability.

In the open-boundary stacked model shown in Figs.~\ref{fig: dwave: Jp_dependence_success_probability}~(a)-(c), antiferromagnetic interactions also enhance the success probability relative to the classical model. 
However, for $J_\mathrm{p}=-1$ and $J_\mathrm{p}=-0.1$ at long annealing time in Figs.~\ref{fig: dwave: Jp_dependence_success_probability}~(a) and (b), the success probability decreases.
Consequently, strong interaction strengths tend to decrease the success probability.
Ferromagnetic interactions consistently degrade the performance, as shown in Figs.~\ref{fig: dwave: Jp_dependence_success_probability}~(d)-(f).
These findings demonstrate that a higher energy scale leads to a deterioration in solution quality.
Furthermore, the distinction between stacked models arises solely from the boundary conditions. 
This suggests that the boundary conditions influence the mechanism that leads to higher success probabilities.
Since the effect of boundary conditions persists across problem sizes $N=101, 201, 301, 401, 501, 601$, the mechanism does scale, as confirmed in Appendix~\ref{appendix: D large spin with small layer}.

In the penalty spin model, the improvement compared to the unprotected model occurs only within a narrower range than in the stacked models.
The effect of the additional interactions on the replicas is essentially the same, independent of their sign.
The distinct behaviors of ferromagnetic and antiferromagnetic interactions arise from differences in penalty-spin alignment during decoding: with antiferromagnetic interactions, penalty spins align opposite to the replicas, whereas with ferromagnetic interactions, they align in the same direction. 
This alignment difference directly leads to the observed disparity in success probability. 
Figure~\ref{fig: dwave: Jp_dependence_success_probability} shows that the penalty spin model with antiferromagnetic interactions sometimes enhances the performance compared to the unprotected model.
The pronounced nature of this effect suggests that, under the present problem setting, penalty spins frequently correspond to the correct solution state after decoding.

These observations can be explained by the mechanism of replica diversification. 
In the stacked model, antiferromagnetic interactions align spins of different replicas in opposite directions, broadening access to diverse low-energy states. 
By contrast, ferromagnetic interactions promote uniform spin alignment across replicas, reducing the diversity of sampled states.

A key factor underlying these behaviors is whether frustration is present among replicas. 
In the periodic-boundary stacked model, frustration arises only when the number of replicas $K$ is odd, allowing access to a broader set of low-energy states. 
This explains why high success probabilities persist even for large negative $J_\mathrm{p}$.
In all cases, appropriately tuned antiferromagnetic interactions yield higher success probabilities than the classical model within shorter annealing times.
Moreover, the periodic-boundary stacked model requires only minimal tuning, making it easier to apply.

\begin{figure*}[t]
    \centering
    \includegraphics[clip,scale=1]{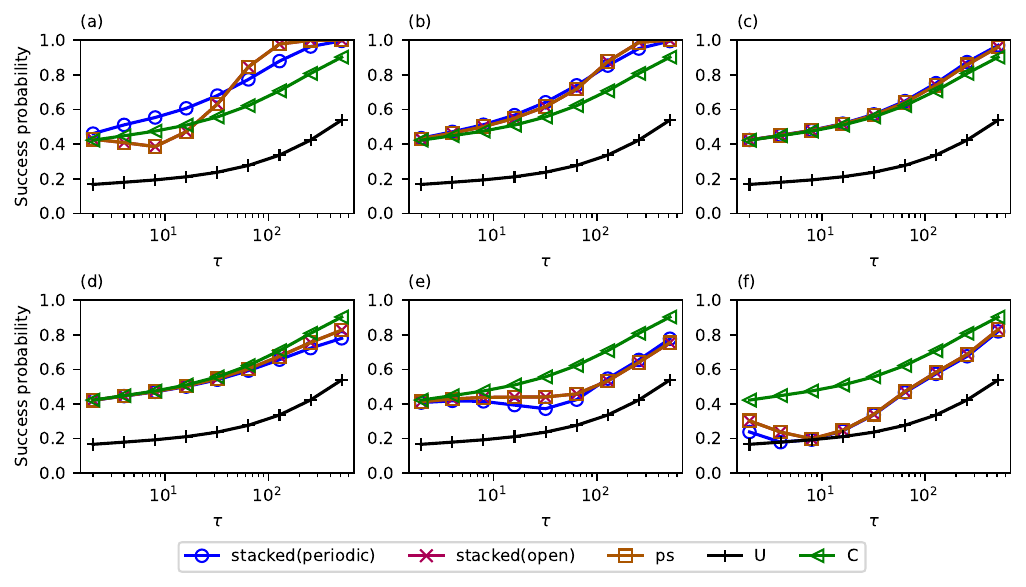}
    \caption{Final success probability as a function of annealing time $\tau$.
    The horizontal axis is shown on a logarithmic scale.
    The problem size is $N = 3$.
    The number of replicas is $K=3$.
    The black and green lines represent the success probabilities of the unprotected and classical models, respectively.
    Each panel shows the results with the different interaction strength $J_\mathrm{p}$: (a) $J_\mathrm{p}=-1$, (b) $J_\mathrm{p}=-0.1$, (c) $J_\mathrm{p}=-0.01$, (d) $J_\mathrm{p}=0.01$, (e) $J_\mathrm{p}=0.1$ and (f) $J_\mathrm{p}=1$.
    Results for the classical model are derived from those of the unprotected model.
    Solid lines between points are provided as a guide to the eye.}
    \label{fig: Jp_dependence_success_probability}
\end{figure*}
\begin{figure*}[t]
    \centering
    \includegraphics[clip,scale=1]{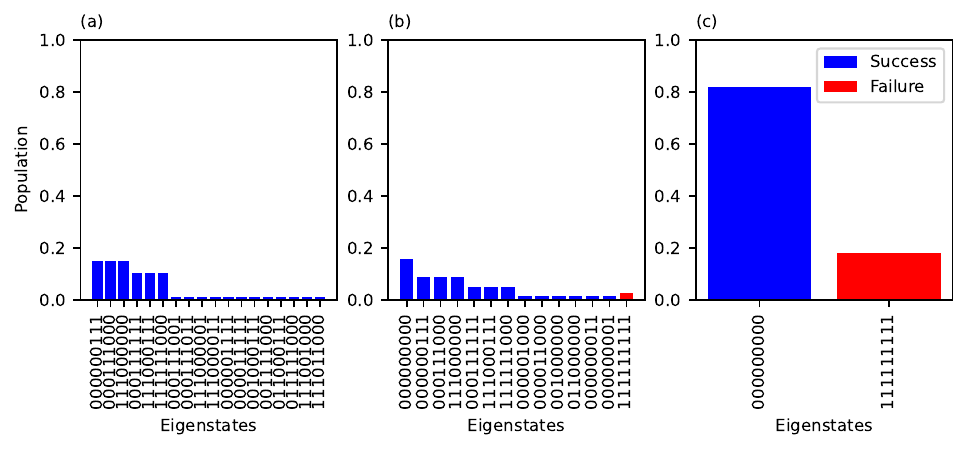}
    \caption{
    Final population in the periodic-boundary stacked model.
    The problem size is $N=3$, with the number of replicas fixed at $K=3$.
    The color bars represent the success probability.
    The red bar represents the failure probability. 
    The annealing time is $\tau = 512$.
    Panels show results for different values of the inter-replica interaction $J_\mathrm{p}$: (a)~$J_\mathrm{p}=-1$, (b)~$J_\mathrm{p}=0$ (classical model) and (c)~$J_\mathrm{p}=1$.
    The eigenstates with probability more than $0.01$ are shown.
    In the bitstring representation, $0$ corresponds to the up state ($\sigma_{i,k}^z=+1$), and $1$ corresponds to the down state ($\sigma_{i,k}^z=-1$).
    The bit at position $l$ is given by $b_l=(1-\sigma_{i, k}^z)/2$, with the rightmost bit $b_0$.}
    \label{fig: population}
\end{figure*}

When frustration is absent, the dominant effect of large interaction strengths is the rescaling of the energy spectrum. 
Previous studies have shown that excessive energy scales degrade solution quality~\cite{raymond2025quantum, hattori2025impact}.
In the open-boundary stacked model and the penalty spin model, where frustration is absent, this rescaling effect governs the performance, resulting in lower success probabilities compared with the periodic-boundary stacked model. 
The influence of energy scaling is also evident in the annealing-time dependence: larger energy scales require longer annealing times to reliably reach low-energy states.
By contrast, in the periodic-boundary stacked model, a high fraction of the diversified low-energy states coincide with the target ground state, enabling robust success probabilities even at short annealing times.
We note, however, that stronger interactions in the periodic-boundary stacked model increase the total number of spins, which may introduce additional errors due to chain breaks; see Appendix~\ref{appendix: influence_of_chain} for details.

To further explore the role of frustration, we next examined the periodic-boundary stacked model.
Unlike the open boundary case, the periodic boundary condition introduces replica frustration, which significantly alters the behavior of the system.
We examined the dependence of success probability on the number of replicas $K$.

Figure~\ref{fig: dwave: success_probability_K} shows the final success probability as a function of the number of replicas $K$ at $\tau=20~\mathrm{\mu s}$ for the frustrated ring model after removing degeneracies with $N=61$.
These results are consistent with those obtained at short annealing times, as shown in Appendix~\ref{appendix: short_annealing_time}.
We emphasize that even the shortest hardware annealing time considered here, $\tau=1~\mu\mathrm{s}$, lies outside the coherent fast annealing regime reported for the one-dimensional Ising chain benchmark in Ref.~\cite{King_2022}. 
Accordingly, the hardware success probabilities, including the unprotected model baseline, may contain thermal and other open system contributions and should not be interpreted as signatures of coherent closed system dynamics.

With antiferromagnetic interactions of moderate strength, increasing $K$ systematically improves the success probability. 
By contrast, with ferromagnetic interactions, no such trend is observed.
Notably, for large antiferromagnetic interaction strength, the success probability exhibited a strong even-odd dependence on $K$, further confirming the importance of frustration in governing performance.

Finally, we analyzed how success probability scales with problem size $N$.
Figure~\ref{fig: dwave: success_probability_scaling} shows the scaling of the final success probability for the frustrated ring model with $N$ after removing degeneracies with $\tau=20~\mathrm{\mu s}$ for $K=15$.
As expected, the success probability decreases with increasing $N$ when $J_\mathrm{p}\geq0.0$.
However, the rate of decrease is significantly mitigated by parallelization: the periodic-boundary stacked model in particular reduces the decay.
Interestingly, 
for some antiferromagnetic interaction strengths, the success probability remains high or increases with $N$.
This behavior can be attributed to the fact that, as the problem size increases, the diversity of the sampled states becomes richer.
Decoding from a more diverse set of samples increases the likelihood of recovering the optimal solution, leading to a higher success probability.
This trend persists across different system sizes and periodic-boundary stacked models, suggesting that the mechanism is robust in periodic-boundary stacked models that utilize replicas in Appendix~\ref{appendix: D large spin with small layer}.

So far, we have presented the results for the frustrated ring model defined in Eq.~\eqref{eq: frustrated_ring_model_without_degenerate}.
These observations are consistent with the frustrated ring model with $h_1=0$.
Results and further discussion of success probabilities are provided in Appendix~\ref{appendix: Frustrated_ring}.

\subsection{Numerical analysis\label{subsec: numerical_analysis}}

\begin{figure*}[t]
    \centering
    \includegraphics[clip,scale=1]{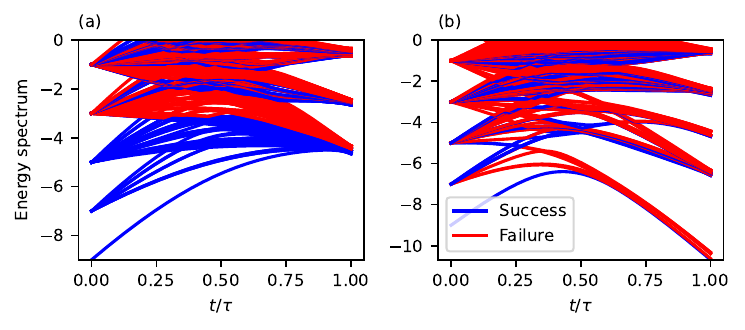}
    \caption{
    Energy spectrum during QA for the periodic-boundary stacked model.
    The problem size is $N=3$, with the number of replicas fixed at $K=3$.
    The blue solid lines represent the energy levels corresponding to the optimal solution after decoding the stacked model, whereas the red solid lines represent those corresponding to the non-optimal solution.
    Panels show results for different values of the inter-replica interaction $J_\mathrm{p}$: (a)~$J_\mathrm{p}=-1$, (b)~$J_\mathrm{p}=1$.
    }
    \label{fig: e_spe}
\end{figure*}
The numerical simulations are intended not only to reproduce the experimental observations but also to clarify their underlying mechanisms.
While the quantum annealer provides success probabilities, the simulations allow analysis of the population distribution and energy spectrum.

Figure~\ref{fig: Jp_dependence_success_probability} shows the final success probability as a function of the annealing time $\tau$ for the frustrated ring model after removing degeneracies with $N=3$ and $K=3$.
The simulations also show that antiferromagnetic interactions enhance the success probability, consistent with the experimental observations.
This indicates that the advantage of antiferromagnetic interactions is not merely a hardware-dependent phenomenon but persists in an ideal, noiseless environment.

In the periodic-boundary stacked model shown in Figs.~\ref{fig: Jp_dependence_success_probability}~(a)-(c), antiferromagnetic interactions enhance the success probability in most cases.
By contrast, in the open-boundary stacked model and the penalty spin model shown in Fig.~\ref{fig: Jp_dependence_success_probability}~(a), the success probability decreases when the annealing time is short and $J_\mathrm{p}=-1$.
These results indicate that energy-scale effects strongly influence success probability and that the replica frustration enhances performance, consistent with the experimental findings.

In the QAC models with ferromagnetic interactions, the success probability decreases compared to the classical model from Figs.~\ref{fig: Jp_dependence_success_probability}~(d)-(f).
Consequently, the QAC models with ferromagnetic interactions are not effective for problems with a small energy gap.

We next analyzed the probability distribution of the sampled states, sorted in ascending order of energy, shown in Fig.~\ref{fig: population}. 
The results demonstrate that in the periodic-boundary stacked model, success states are broadly distributed among the low-energy states. 
Consequently, the correct solution can be obtained without requiring fully adiabatic evolution, explaining why the diversification in this model is effective, and this model yields high success probabilities.

The total spectrum of $K$ independent replicas is not identical to that of a single replica.
Each replica possesses the same single-replica spectrum and the same minimum-gap bottleneck, while the overall dynamics factorize into independent single-replica dynamics. 
Consequently, the decoded success probability increases according to Eq.~\eqref{eq: suc_prob_C} shown in Appendix~\ref{appendix:B probability}.
Therefore, for problem instances intrinsically hard to solve by QA, parallelization provides little benefit. 
In theory, the benefit of parallelization scales with the number of replicas $K$.
A detailed discussion is provided in Appendix~\ref{appendix:B probability}. 
For such hard problems, introducing diversity across replicas is essential for achieving a high success probability.

We also compared the energy spectra obtained by exact diagonalization for different QAC models.
Figure~\ref{fig: e_spe} shows the energy spectrum of the periodic-boundary stacked model for $J_\mathrm{p}=-1$ and $J_\mathrm{p}=1$.
In the periodic-boundary stacked model, success states occupy a wide range of low-energy levels, implying that even diabatic time evolution can readily yield correct solutions.
The antiferromagnetic interactions preserve the correspondence between the low-energy eigenstates and the optimal solution shown in the previous study~\cite{pearson2019analog, bennett2023using}.
Moreover, Fig.~\ref{fig: e_spe}~(a) shows that the success states spread to the low-energy-excited states.
The results indicate the potential to obtain a high success probability even in a shorter annealing time required in the adiabatic regime.
On the other hand, the eigenstates corresponding to the optimal solution spread into higher-energy eigenstates in the periodic-boundary stacked model with ferromagnetic interactions.

In the periodic-boundary stacked model, the decodability of low-energy states varies with different interaction strengths.
For $J_\mathrm{p}=-1$ and $-0.1$, the $47$ lowest-energy states correspond to the optimal solution.
For $J_\mathrm{p}= -0.01$, the $38$ lowest-energy states correspond to the optimal solution.
For $J_\mathrm{p}=-0.001,0, 0.001$, the $6$ lowest-energy states correspond to the optimal solution.
For $J_\mathrm{p}= 0.01,0.1,1$, only the ground state corresponds to the optimal solution.
Since these properties are preserved in Ising machines, the periodic-boundary stacked model with antiferromagnetic interactions may also be beneficial for other ground-state search algorithms.
In an open boundary stacked model with antiferromagnetic interactions, the $2$ lowest energy states correspond to the optimal solution.
The absence of frustration significantly decreases the effects of spreading the success states to low-energy states.
The results indicate that frustration is an important property for corresponding the multiple low-energy states to success states.

By contrast, in the open-boundary stacked model and the penalty spin model, the first- or second-excited states do not correspond to success states.
As a result, adiabatic evolution is necessary to reach the ground state. 
In these QAC models, adding interactions enlarges the energy gap between the ground and first excited states, facilitating adiabatic evolution and improving performance.

Finally, note that in noisy or embedded settings, certain states may become stabilized due to hardware-specific effects. 
Under such conditions, models that inherently diversify sampled solutions, such as the periodic-boundary stacked model, exhibit superior performance. 
This observation is consistent with the high success probabilities observed experimentally on the quantum annealer.

\section{\label{sec: conclusion}Conclusion}
We examined the properties of QAC models with additional interactions in QA characterized by a small energy gap, using the frustrated ring model as a benchmark.
By employing the frustrated ring model, we ensured that the observed performance reduced embedding overhead and separately assessed chain-break effects.
As a result, we demonstrated the potential of QAC models to overcome limitations of conventional QA.
Furthermore, this study provides insights into the mechanisms underlying the performance improvement.

Experiments on the actual quantum annealer and numerical simulations revealed that antiferromagnetic interactions enhance the success probability, whereas ferromagnetic interactions suppress it.
Upon decoding, the QAC models yield various eigenstates corresponding to the optimal solution of the unprotected model.
Notably, incorporating antiferromagnetic interactions within the QAC models significantly increases the success probability, even at short annealing times. 
This suggests that the QAC models are a promising strategy for hardware with limited coherence time and control noise.

We demonstrated that frustration introduced by periodic boundary conditions plays a key role in diversifying low-energy configurations, thereby enabling higher success probabilities even in the short annealing-time regime.
These findings suggest that the QAC models can facilitate diabatic QA under limited annealing times.

These results demonstrate that the proposed method enhances the success probability within shorter annealing times. 
This improvement, however, comes at the expense of requiring a larger number of physical qubits to treat small–energy-gap problems, for which conventional approaches typically demand significantly longer annealing time.
Importantly, while the annealing time required in conventional schemes is generally expected to increase exponentially as the minimum energy gap decreases, our observations do not indicate a corresponding exponential growth in the qubit overhead of the proposed method—at least within the range of system sizes examined in this study.
These findings suggest a shift in the dominant resource scaling from annealing time to qubit count.
In this sense, expanding the number of available qubits may be more impactful than extending the maximum annealing time when addressing small-gap instances.

In the future, our results suggest several promising directions.
First, optimizing the interaction strength relative to the annealing time is essential to balance the benefits of replica frustration against the detrimental effects of large energy scales.
Second, developing hardware-aware QAC models, tailored to specific connectivity and embedding constraints, will be vital for future QA platforms.
Since constraints in combinatorial optimization are typically imposed as penalties, many problems reduce to dense Ising models. 
For practical applications, it is crucial to develop approaches that minimize embedding overhead and clarify how embedding affects performance.
Third, a systematic investigation of the qualitative differences between simulated quantum annealing~\cite{santoro2002theory} and open-system dynamics described by the Redfield master equation~\cite{redfield1957theory} is crucial for bridging the gap between numerical simulations and the behavior of realistic quantum annealing hardware.
Finally, extending the present framework to more general optimization problems will further elucidate the role of frustration and interaction tuning in achieving scalability in quantum annealers.

\begin{acknowledgments}
The authors would like to thank Kanta Hino and Tetsuro Abe for helpful discussions and valuable comments.
This work was partially supported by the Japan Society for the Promotion of Science (JSPS) KAKENHI (Grant Number JP23H05447), the Council for Science, Technology, and Innovation (CSTI) through the Cross-ministerial Strategic Innovation Promotion Program (SIP), ``Promoting the application of advanced quantum technology platforms to social issues'' (Funding agency: QST), Japan Science and Technology Agency (JST) (Grant Number JPMJPF2221). 
The computations in this work were partially performed using the facilities of the Supercomputer Center, the Institute for Solid State Physics, The University of Tokyo.
S.~Tanaka wishes to express their gratitude to the World Premier International Research Center Initiative (WPI), MEXT, Japan, for their support of the Human Biology-Microbiome-Quantum Research Center (Bio2Q).
T.~H. was supported by JST SPRING (Grant Number JPMJSP2123).
\end{acknowledgments}

\appendix
\section{Effects of embedding\label{appendix: influence_of_chain}}
In this section, we examined the effects of constructing embedding chains in a quantum annealer.
We performed the embeddings with minor embedding~\cite{Choi_2008, Choi_2011}.
Ideally, all spins within the chain should take the same value.
However, due to noise, insufficient chain strength, or conflicting problem interactions, some chains may break, causing spins within the same chain to disagree.
\begin{figure}[t]
    \centering
    \includegraphics[clip,scale=1.0]{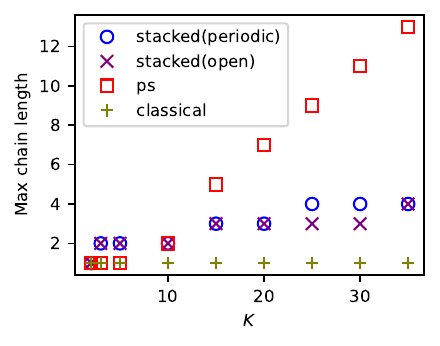}
    \caption{The maximum chain length as a function of the number of replicas $K$ for different QAC models. 
    The problem size is $N=61$.
    }
    \label{fig: chain_length}
\end{figure}
\begin{figure*}[t]
    \centering
    \includegraphics[clip,scale=1.0]{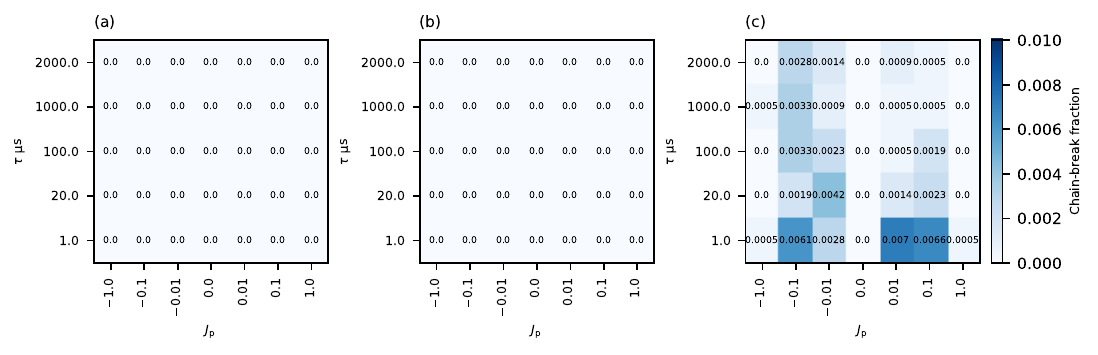}
    \caption{The chain-break fractions as a function of the interaction strength $J_\mathrm{p}$ and the annealing time $\tau~\mathrm{\mu s}$ for different QAC models.
    The problem size is $N=61$, with the number of replicas fixed at $K=35$.
    Panels show (a)~periodic-boundary stacked model, (b)~open-boundary stacked model, and (c)~penalty spin model.
    The results at $J_\mathrm{p}=0$ correspond to the classical model with $K=35$.
    }
    \label{fig: chain_fraction}
\end{figure*}
Figure~\ref{fig: chain_length} shows the maximum chain length for each QAC model.
Under the present problem setting, the penalty spin model requires the longest chain length.
This occurs because, as the number of replicas increases, the number of interactions involving the penalty spins also grows.
Longer chains are more susceptible to chain breaks, and avoiding them requires setting the chain strength, the interactions between spins in a chain, to large values. 
However, large chain strengths constrain the overall energy scale of the problem, reducing the solution quality.
In the penalty spin model, we also observed greater variability in chain lengths.
The fractions of penalty spins effectively diminish their role, suggesting that the stacked model becomes more advantageous as the underlying problem becomes denser.

Figure~\ref{fig: chain_fraction} shows the chain-break fraction for each QAC model.
Ideally, the spins in a chain are aligned in the same direction.
The chain-break fraction is the ratio of the number of misaligned spins to the total number of spins in a chain.
This metric serves as an indicator of embedding overhead and correlates with the susceptibility of chain breaks.

The results indicate that the chain-break fraction is significantly larger in the penalty spin model shown in Fig.~\ref{fig: chain_fraction}~(c) than the open-boundary and periodic-boundary stacked models shown in Figs.~\ref{fig: chain_fraction}~(a) and (b). 
When $K=15$, the chain-break fraction is all zero in all QAC models.
Our results demonstrate that the stacked model is more embedding-friendly. 
The reason is that, in the penalty spin model, the maximum number of interactions per spin grows as the sum of $(K-1)$ along the parallelization directions and the number of interactions in the unprotected model.
In the stacked model, by contrast, the contribution from parallelization is limited to two, regardless of $K$. 
Consequently, for $K \geq 3$, the interaction density is lower in the stacked model than in the penalty spin model.

These results suggest that certain QAC models are inherently better suited to the topology of specific hardware architectures, and that the development of QAC models tailored to future hardware topologies will be essential for minimizing embedding overhead.

\section{Selection of the efficient number of replicas\label{appendix: TTS}}
 The efficient number of replicas is selected by the TTS considering the qubit number efficiency.
The metric is defined as
\begin{align}
    \label{eq: TTS_K}
    T_K=\tau \frac{\log(1-0.99)}{\log(1-p(\tau))}K.
\end{align}
The metric considers the efficiency of the qubit by multiplying the number of replicas.
By using $T_K$, we evaluate the efficient number of replicas.

\begin{figure}
    \centering
    \includegraphics[clip,scale=1.0]{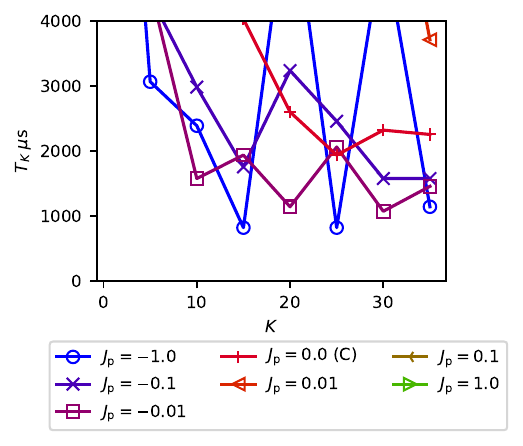}
    \caption{$T_K$ as a function of the number of replicas $K$ with the periodic-boundary stacked model.
    The problem size is $N = 61$.
    The annealing time is $\tau=20~\mathrm{\mu s}$.
    The results at $J_\mathrm{p}=0$ correspond to the classical model.
    Solid lines between points are provided as a guide to the eye.
    The jumps in $T_K$ are associated with an even–odd effect in $K$. 
    For antiferromagnetic inter-replica couplings with periodic boundary conditions, odd $K$ produces a frustrated replica cycle, whereas all inter-replica couplings can be simultaneously satisfied for even $K$.
    }
    \label{fig: dwave: TTS_K}
\end{figure}

Figure~\ref{fig: dwave: TTS_K} shows $T_K$ as a function of $K$ on an actual quantum annealer.
According to the $T_K$ metric, the optimal efficiency is achieved at $K=15$.
Notably, for odd values of $K$, strong antiferromagnetic interactions~($J_\mathrm{p}=-1$) exhibit a clear performance enhancement.
The nonmonotonic changes in $T_K$ are associated with the parity of $K$.
In the periodic-boundary stacked model with antiferromagnetic inter-replica interactions, an even number of replicas permits an alternating configuration that satisfies all couplings along the replica direction. 
For odd $K$, however, the periodic replica direction forms an odd antiferromagnetic cycle, and at least one coupling must remain unsatisfied.
This parity-induced frustration modifies the low-energy spectrum and the effectiveness of obtaining the success states.
The jumps in $T_K$ between neighboring values of $K$ should therefore be understood as an even–odd effect rather than as statistical fluctuations or a smooth dependence on $K$.
Although the present results indicate an even–odd dependence, a more systematic investigation using a finer sampling of $K$ values is required to fully characterize this parity effect, which we leave for future work.

\section{Comparison of two independent $K=3$ QAC models and a single $K=6$ QAC model\label{appendix:parallel_QAC}}
\begin{figure}[t]
    \centering
    \includegraphics[clip,scale=1.0]{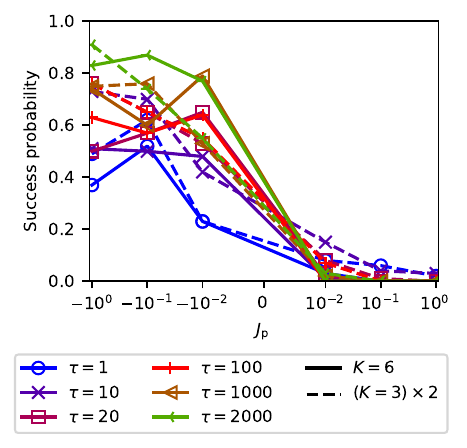}
    \caption{
    The final success probability as a function of the interaction strength $J_\mathrm{p}$ with the periodic-boundary stacked model.
    The problem size is $N = 61$.
    A symmetric logarithmic scale is used for the horizontal axis to capture variations around zero.
    Solid lines represent the periodic-boundary stacked model with $K=6$.
    Dotted lines represent the two independent periodic-boundary stacked models with $K=3$.
    Solid and dotted lines between points are provided as a guide to the eye.
    }
    \label{fig: dwave: two_K=3andK=6}
\end{figure}

From the perspective of exploiting the qubit resource, comparisons between the parallel QAC models and the QAC models are important.
In this appendix, we compare the two independent $K=3$ QAC models and a single $K=6$ QAC model.
Figure~\ref{fig: dwave: two_K=3andK=6} shows the final success probability comparison between the two independent $K=3$ periodic-boundary stacked models and the single $K=6$ periodic-boundary QAC model.
Note that these results were obtained by \verb|Advantage2_system1.13|.
As shown in Fig.~\ref{fig: dwave: two_K=3andK=6}, the two parallelized $K=3$ periodic-boundary stacked models and the single $K=6$ stacked model exhibit nearly identical success probability behavior.

The slight differences observed can be attributed to the presence or absence of inter-replica interactions.
In the $K=6$ model, all replicas are coupled, whereas in the two independent $K=3$ models, they are not.
As $K$ increases, the distinction between parallelized and single QAC implementations is expected to become negligible.
In the present case, where $K$ is relatively small ($K=3$ and $K=6$), individual inter-replica couplings can have a noticeable impact.
Nevertheless, we find that their overall effect remains limited. 
Thus, as the system size increases, the differences become negligible.
While further investigation is required for larger system sizes or in the presence of significant embedding overhead, we leave this for future work.

\section{Success probability in a short annealing time\label{appendix: short_annealing_time}}
\begin{figure}[t]
    \centering
    \includegraphics[clip,scale=1.0]{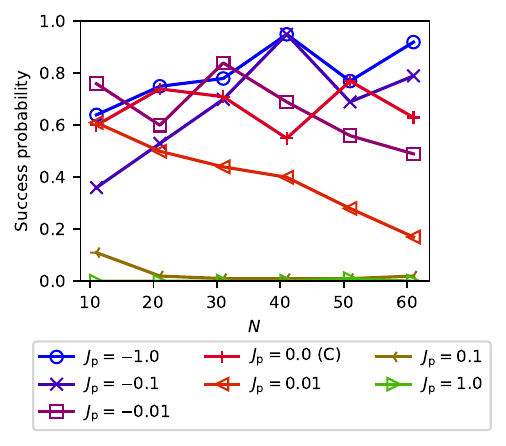}
    \caption{
    Scaling of the final success probability of the periodic-boundary stacked model.
    The number of replicas is fixed to $K=15$.
    The annealing time is $\tau=1~\mathrm{\mu s}$.
    The results at $J_\mathrm{p}=0$ correspond to the classical model.
    The dashed black line represents the result of the success probability with the unprotected model.
    Solid lines between points are provided as a guide to the eye.
    }
    \label{fig: dwave: tau1_success_probability_scaling}
\end{figure}
\begin{figure}[t]
    \centering
    \includegraphics[clip,scale=1.0]{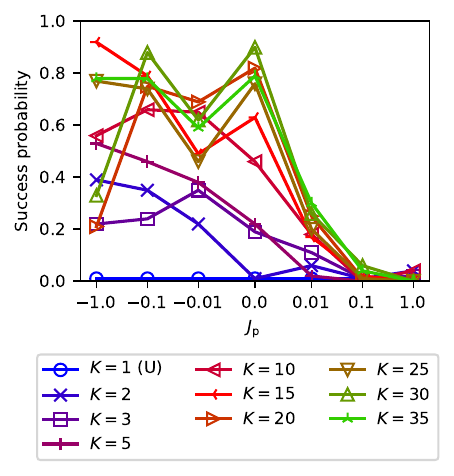}
    \caption{
    The final success probability as a function of the interaction strength $J_\mathrm{p}$ with the periodic-boundary stacked model.
    The problem size is $N = 61$.
    The annealing time is $\tau=1~\mathrm{\mu s}$.
    The results at $J_\mathrm{p}=0$ correspond to the classical model.
    A symmetric logarithmic scale is used for the horizontal axis to capture variations around zero.
    The results at $K=1$ show the results of the unprotected model.
    Solid lines between points are provided as a guide to the eye.
    }
    \label{fig: dwave: tau1_success_probability_Jp}
\end{figure}
In this section, we examined the size scaling and parameter dependence of the QAC models in a short annealing-time regime.
Although the previous figures presented results for $\tau = 20~\mathrm{\mu s}$, here we focus on the case of $\tau = 1~\mathrm{\mu s}$. 
Figure~\ref{fig: dwave: tau1_success_probability_scaling} shows the size dependence of the success probability at $\tau = 1~\mathrm{\mu s}$.

When negative $J_\mathrm{p}$ is introduced, the degradation with system size is significantly improved, and the success probability becomes higher than that of the classical model.
In the cases $J_\mathrm{p}=-0.1,-1$, the success probability increases as the system size increases since the diversity of the sampled state increases.
This demonstrates that antiferromagnetic interactions provide a clear advantage even in the short-time regime, consistent with the results obtained for long annealing times.

Moreover, the success probabilities achieved by the QAC models at $\tau = 1~\mathrm{\mu s}$ are comparable to those of the classical model with annealing times in the range $\tau = 100$–$1000~\mathrm{\mu s}$. 
This result highlights the effectiveness of QAC models in environments constrained by coherence times and noise. 
The results further indicate that leveraging a larger number of qubits can mitigate noise effects, suggesting that QAC models may serve as practical error-mitigation strategies for noisy quantum computers.

Figure~\ref{fig: dwave: tau1_success_probability_Jp} shows the final success probability as a function of the strength of the additional interaction.
The results confirm once again that parallelization with antiferromagnetic interactions yields higher success probabilities.
In the classical model, the success probability increases monotonically as the number of replicas increases.
In contrast, in the periodic-boundary stacked model, increasing the number of replicas does not necessarily lead to a corresponding increase in success probability.
We attribute this to the stronger impact of the energy scale at short annealing times.
As the annealing time decreases, the relative effect of large interaction strengths becomes increasingly pronounced.
This finding also implies that even in the periodic-boundary stacked model, where frustration enhances performance, the energy-scale effect can reduce the success probability. 
As the problem size and number of replicas increase, the number of interactions grows, making the energy-scale effect even more dominant.

In the periodic-boundary stacked model, successful solutions are mapped onto a set of low-energy eigenstates.
However, if the quantum state transitions to higher excited states outside these low-energy states, the decoding process becomes less effective, and the success probability does not improve significantly.
Consequently, the enhancement in success probability achieved by the periodic-boundary stacked model is smaller than that obtained by the classical model when simply increasing the number of replicas.
The periodic-boundary stacked model is therefore most effective in regimes where the population is dominated by the ground state, the first excited state, or nearby low-energy excited states.
Such behavior is commonly observed in systems where the energy gap decreases exponentially with system size, and the periodic-boundary stacked model is therefore particularly effective.
In such regimes, decoding can recover a large fraction of successful solutions, and the antiferromagnetic periodic-boundary stacked model exhibits its strongest performance.

\section{Results for the larger Ising model\label{appendix: D large spin with small layer}}
The performance of the periodic-boundary stacked models in the larger Ising model was examined.
We set the number of spins in the unprotected model $N=101, 201, 301, 401, 501, 601$ and the number of replicas to $K=1,2,3,4,5$, considering the overhead arising from differences in the boundary conditions between the Zephyr graph and the QAC models.
The annealing time was set to $\tau = 1, 20, 100, 1000, 2000~\mathrm{\mu s}$.
We examined the dependence of performance on annealing time, additional interactions in the QAC models, and the number of spins.
The quantum annealer was executed with 100 runs, and embedding with minimal overhead was selected for 1000 embeddings with minor embedding.
All other parameters were set to their default values.

\begin{figure}
    \centering
    \includegraphics[scale=1.0,clip]{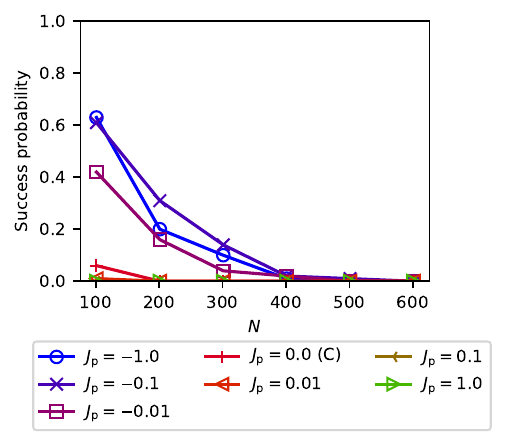}
    \caption{Scaling of the final success probability of the periodic-boundary stacked model.
    The number of replicas is fixed to $K=5$.
    The annealing time is $\tau=20~\mathrm{\mu s}$.
    The results at $J_\mathrm{p}=0$ correspond to the classical model.
    The dashed black line represents the result of the success probability with the unprotected model.
    Solid lines between points are provided as a guide to the eye.}
    \label{fig: dwave N_vs_success_probability_large_spin_tau20}
\end{figure}

Figure~\ref{fig: dwave N_vs_success_probability_large_spin_tau20} shows the final success probability as a function of $N$.
 As the system size increases, solving the problem becomes increasingly difficult for the classical model.
However, in the antiferromagnetic periodic-boundary stacked model, although the success probability decreases with system size, successful solutions can still be obtained.
These results indicate that the proposed approach is effective for problem instances that are difficult to solve using conventional QA and are also challenging for classical models.

\begin{figure}
    \centering
    \includegraphics[scale=1.0,clip]{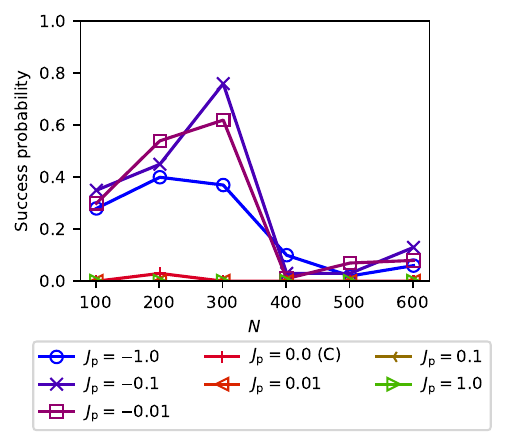}
    \caption{Scaling of the final success probability of the periodic-boundary stacked model.
    The number of replicas is fixed to $K=3$.
    The annealing time is $\tau=1000~\mathrm{\mu s}$.
    The results at $J_\mathrm{p}=0$ correspond to the classical model.
    The dashed black line represents the result of the success probability with the unprotected model.
    Solid lines between points are provided as a guide to the eye.}
    \label{fig: dwave N_vs_success_probability_large_spin_tau1000}
\end{figure}

From Fig.~\ref{fig: dwave N_vs_success_probability_large_spin_tau1000}, when antiferromagnetic interactions are introduced, the success probability increases with the system size $N$ and decreases at $N=301$.
This behavior can be attributed to the fact that, as the problem size increases, the diversity of the sampled states becomes richer.
In the periodic-boundary stacked model, successful solutions are mapped onto a set of low-energy eigenstates.
However, if the quantum state transitions to higher excited states outside these low-energy states, the decoding process becomes less effective, and the success probability does not improve significantly.
The periodic-boundary stacked model is therefore most effective in regimes where the population is dominated by the ground state, the first excited state, or nearby low-energy excited states.
Such behavior is commonly observed in systems where the energy gap decreases exponentially with system size, and the periodic-boundary stacked model is therefore particularly effective.
The changing points of the trade-offs between the energy scale and the diversity effects are about $N=301$ in this model.
From these results, the periodic-boundary stacked model is effective for even larger Ising models.

\section{Success probability in a frustrated ring model\label{appendix: Frustrated_ring}}
\begin{figure}[t]
    \centering
    \includegraphics[clip,scale=1.0]{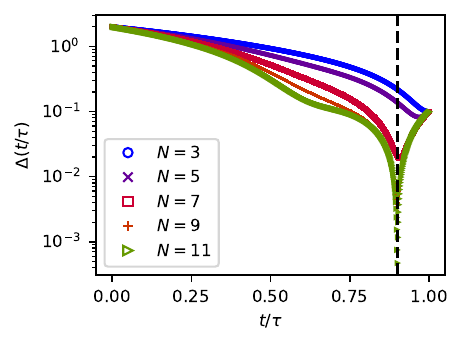}
    \caption{Energy gap $\Delta(t/\tau)$ between the ground state and the first excited state during QA for the frustrated ring model.
    The interaction parameters are $J=1$, $J_\mathrm{L}=0.5$ and $J_\mathrm{R}=0.45$.
    The black dashed line indicates the time at which the energy gap closes at $N\rightarrow\infty$, obtained analytically from the frustrated ring model with $h_1=0$ in Eq.~\eqref{eq: frustrated_ring_model_without_degenerate}.
    }
    \label{fig: energy_gap_FRmodel_with_degenerate}
\end{figure}
\begin{figure*}[t]
    \includegraphics[clip,scale=1.0]{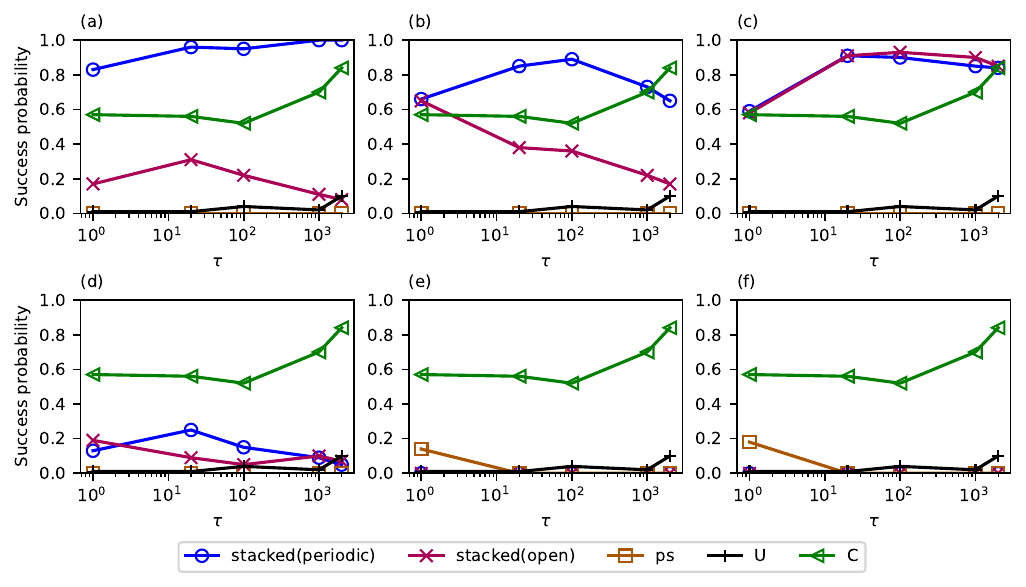}
        \caption{Final success probability as a function of annealing time $\tau$ for the frustrated ring model with $h_1=0$.
    The horizontal axis is shown on a logarithmic scale.
    The problem size is $N = 61$.
    The number of replicas is $K=15$.
    The black and green lines represent the success probabilities of the unprotected and classical models, respectively.
    Each panel shows the results with the different interaction strength $J_\mathrm{p}$: (a) $J_\mathrm{p}=-1$, (b) $J_\mathrm{p}=-0.1$, (c) $J_\mathrm{p}=-0.01$, (d) $J_\mathrm{p}=0.01$, (e) $J_\mathrm{p}=0.1$ and (f) $J_\mathrm{p}=1$.
    Solid lines between points are provided as a guide to the eye.
    }
    \label{fig: dwave: Jp_dependence_success_probability_frustrated_ring}
\end{figure*}
In this section, we present the success probabilities of the QAC models applied to the frustrated ring model defined without local magnetic fields ($h_1=0$ in Eq.~\eqref{eq: frustrated_ring_model_without_degenerate}).
The energy gap of this model has been analytically studied in previous works, and it is known to close exponentially in the limit of infinite system size $N$.
Figure~\ref{fig: energy_gap_FRmodel_with_degenerate} shows the energy gap of the frustrated ring model, obtained by exact diagonalization. 
As $N$ increases, the energy gap closes, the ground state becomes degenerate between the all-up and all-down spin configurations, and the first excited state is fourfold degenerate. 

Figure~\ref{fig: dwave: Jp_dependence_success_probability_frustrated_ring} shows the final success probability as a function of annealing time $\tau$. 
For the unprotected model, the success probability remains below $0.1$ in most cases, regardless of the annealing time.
The settings of the actual quantum annealer are the same as those in Section~\ref{sec: settings}.
Consistent with the results in Section~\ref {sec: result}, high success probabilities are obtained with antiferromagnetic interactions. 
In particular, for the periodic-boundary stacked model, the success probability is enhanced over a broader parameter range than when degeneracy is lifted. 
This improvement is attributed to the fourfold degeneracy of the first excited state, which provides a diversity of accessible states and enables high success probabilities even at shorter annealing times. 
Thus, for degenerate problems, the diversity of solutions enhances the likelihood that the excited states of the parallelized model correspond to the correct solutions.
This property confirms the effectiveness of QAC models even in degenerate systems.

The open-boundary stacked model also exhibits a wide parameter region with high success probabilities, indicating that QAC models are effective even without frustration.
However, for large absolute values of the interaction strength, the success probability decreases, which we attribute to energy-scale effects. 
Therefore, the alleviation of fine-tuning requirements by frustration is consistent with our interpretation.

The penalty spin model exhibits qualitatively similar behavior. 
Adding interactions of small absolute magnitude increases the success probability compared to the unprotected model; however, the success probability of the penalty spin model is lower than that of the other QAC models.

Taken together, these results demonstrate that for problems otherwise hard to solve with quantum annealing, spatial parallelization combined with additional interactions provides a viable strategy. 
This highlights the potential of QAC models as near-term algorithms applicable to a wide range of problems.

\section{Probability of the classical model\label{appendix:B probability}}
The classical model facilitates obtaining the optimal solution through decoding.
In this section, we discuss the probabilities associated with the classical model.
When the ground state is non-degenerate, the number $n_\mathrm{suc}$ of success states in the classical model with $N$ spins and $K$ replicas is given by
\begin{align}
    \label{eq: number_of_success}
    n_\mathrm{suc}&=\sum_{k=1}^K \binom{K}{K-k+1}(2^N-1)^{k-1}\nonumber\\
    &=2^{NK}-(2^N-1)^K.
\end{align}
Here, each replica is independent in the classical model.

If the success probability of the unprotected model is denoted by $p_\mathrm{success}^\mathrm{U}$, the success probability of the classical model is given by
\begin{align}
    \label{eq: suc_prob_C}
    p_\mathrm{success}^\mathrm{C}= 1-\left(1-p_\mathrm{success}^\mathrm{U}\right)^K.
\end{align}
The ground-state probability of the classical model is given by
\begin{align}
    \label{eq: gs_prob_C}
    p_\mathrm{gs}^\mathrm{C}= \left(p_\mathrm{gs}^\mathrm{U}\right)^K.
\end{align}


\bibliography{reference}

@article{kadowaki1998quantum,
  title={Quantum annealing in the transverse {I}sing model},
  author={Kadowaki, T. and Nishimori, H.},
  journal={Phys. Rev. E},
  volume={58},
  number={5},
  pages={5355},
  year={1998},
  publisher={APS}
}

@article{farhi2000quantum,
  title={Quantum computation by adiabatic evolution},
  author={Farhi, E. and Goldstone, J. and Gutmann, S. and Sipser, M.},
  journal={arXiv preprint quant-ph/0001106},
  year={2000}
}

@book{mcgeoch2014adiabatic,
  title={Adiabatic quantum computation and quantum annealing: {T}heory and practice},
  author={McGeoch, C. C.},
  year={2014},
  publisher={Morgan \& Claypool Publishers}
}

@book{tanaka2017quantum,
  title={Quantum spin glasses, annealing and computation},
  author={Tanaka, S. and Tamura, R. and Chakrabarti, B. K.},
  year={2017},
  publisher={Cambridge University Press}
}

@article{albash2018adiabatic,
  title={Adiabatic quantum computation},
  author={Albash, T. and Lidar, D. A.},
  journal={Rev. Mod. Phys.},
  volume={90},
  number={1},
  pages={015002},
  year={2018},
  publisher={APS}
}

@article{chakrabarti2023quantum,
  title={Quantum annealing and computation: challenges and perspectives},
  author={Chakrabarti, Bikas K and Leschke, Hajo and Ray, Purusattam and Shirai, Tatsuhiko and Tanaka, Shu},
  journal={Philos. Trans. R. Soc. A},
  volume={381},
  number={2241},
  pages={20210419},
  year={2023},
  publisher={The Royal Society}
}

@article{kato1950adiabatic,
  title={On the adiabatic theorem of quantum mechanics},
  author={Kato, T.},
  journal={J. Phys. Soc. Jpn.},
  volume={5},
  number={6},
  pages={435--439},
  year={1950},
  publisher={The Physical Society of Japan}
}

@article{QuTiP1,
    author = {Johansson, J.R. and  Nation,P. D.  and Nori,F.},
    title = {Qu{T}i{P}: An open-source {P}ython framework for the dynamics of open quantum systems.},
    year = {2012},
    journal = {Comput. Phys. Commun.},
    volume = {183},
    issue = {8},
    pages = {1760}
    }

@article{QuTiP2,
    author = {Johansson,J.R. and  Nation,P. D.  and Nori,F.},
    title = {Qu{T}i{P} 2: A {P}ython framework for the dynamics of open quantum systems.},
    year = {2013},
    journal = {Comput. Phys. Commun.},
    volume = {184},
    issue = {4},
    pages = {1234}
    }

@article{lucas2014ising,
  title={Ising formulations of many {NP} problems},
  author={Lucas, Andrew},
  journal={Front. Phys.},
  volume={2},
  pages={5},
  year={2014},
  publisher={Frontiers Media SA}
}

@article{tanahashi2019application,
  title={Application of {I}sing machines and a software development for {I}sing machines},
  author={Tanahashi, Kotaro and Takayanagi, Shinichi and Motohashi, Tomomitsu and Tanaka, Shu},
  journal={J. Phys. Soc. Jpn.},
  volume={88},
  number={6},
  pages={061010},
  year={2019},
  publisher={The Physical Society of Japan}
}

@article{crosson2021prospects,
  title={Prospects for quantum enhancement with diabatic quantum annealing},
  author={Crosson, EJ and Lidar, DA},
  journal={Nat. Rev. Phys.},
  volume={3},
  number={7},
  pages={466--489},
  year={2021},
  publisher={Nature Publishing Group UK London}
}

@article{cote2023diabatic,
  title={Diabatic quantum annealing for the frustrated ring model},
  author={C{\^o}t{\'e}, J. and Sauvage, F. and Larocca, M. and Jonsson, M. and Cincio, L. and Albash, T.},
  journal={Quantum Sci. and Technol.},
  volume={8},
  number={4},
  pages={045033},
  year={2023},
  publisher={IOP Publishing}
}

@article{feinstein2025robustness,
  title={Robustness of diabatic enhancement in quantum annealing},
  author={Feinstein, Natasha and Shalashilin, Ivan and Bose, Sougato and Warburton, PA},
  journal={Quantum Sci. Technol.},
  volume={10},
  number={2},
  pages={025011},
  year={2025},
  publisher={IOP Publishing}
}

@article{FMA2020,
  title = {Designing metamaterials with quantum annealing and factorization machines},
  author = {Kitai, Koki and Guo, Jiang and Ju, Shenghong and Tanaka, Shu and Tsuda, Koji and Shiomi, Junichiro and Tamura, Ryo},
  journal = {Phys. Rev. Res.},
  volume = {2},
  issue = {1},
  pages = {013319},
  numpages = {10},
  year = {2020},
  month = {Mar},
  publisher = {American Physical Society},
  doi = {10.1103/PhysRevResearch.2.013319},
  url = {https://link.aps.org/doi/10.1103/PhysRevResearch.2.013319}
}

@article{FMA2022,
  title = {Continuous black-box optimization with an {I}sing machine and random subspace coding},
  author = {Izawa, Syun and Kitai, Koki and Tanaka, Shu and Tamura, Ryo and Tsuda, Koji},
  journal = {Phys. Rev. Res.},
  volume = {4},
  issue = {2},
  pages = {023062},
  numpages = {9},
  year = {2022},
  month = {Apr},
  publisher = {American Physical Society},
  doi = {10.1103/PhysRevResearch.4.023062},
  url = {https://link.aps.org/doi/10.1103/PhysRevResearch.4.023062}
}

@article{photonic_laser2022,
	author = {Takuya Inoue and Yuya Seki and Shu Tanaka and Nozomu Togawa and Kenji Ishizaki and Susumu Noda},
	doi = {10.1364/OE.476839},
	journal = {Opt. Express},
	month = {Nov},
	number = {24},
	pages = {43503--43512},
	publisher = {Optica Publishing Group},
	title = {Towards optimization of photonic-crystal surface-emitting lasers via quantum annealing},
	url = {https://opg.optica.org/oe/abstract.cfm?URI=oe-30-24-43503},
	volume = {30},
	year = {2022}}

@article{FMA2023,
  title = {Quantum Annealing Optimization Method for the Design of Barrier Materials in Magnetic Tunnel Junctions},
  author = {Nawa, Kenji and Suzuki, Tsuyoshi and Masuda, Keisuke and Tanaka, Shu and Miura, Yoshio},
  journal = {Phys. Rev. Appl.},
  volume = {20},
  issue = {2},
  pages = {024044},
  numpages = {15},
  year = {2023},
  month = {Aug},
  publisher = {American Physical Society},
  doi = {10.1103/PhysRevApplied.20.024044},
  url = {https://link.aps.org/doi/10.1103/PhysRevApplied.20.024044}
}

@article{matsumori2022application,
  title={Application of {QUBO} solver using black-box optimization to structural design for resonance avoidance},
  author={Matsumori, Tadayoshi and Taki, Masato and Kadowaki, Tadashi},
  journal={Sci. Rep.},
  volume={12},
  number={1},
  pages={12143},
  year={2022},
  publisher={Nature Publishing Group UK London}
}

@inproceedings{rosenberg2015solving_trading,
  title={Solving the optimal trading trajectory problem using a quantum annealer},
  author={Rosenberg, Gili and Haghnegahdar, Poya and Goddard, Phil and Carr, Peter and Wu, Kesheng and De Prado, Marcos L{\'o}pez},
  booktitle={Proceedings of the 8th workshop on high performance computational finance},
  pages={1--7},
  year={2015}
}

@article{grant2021benchmarking_portofolio,
  title={Benchmarking quantum annealing controls with portfolio optimization},
  author={Grant, Erica and Humble, Travis S and Stump, Benjamin},
  journal={Phys. Rev. Appl.},
  volume={15},
  number={1},
  pages={014012},
  year={2021},
  publisher={APS}
}

@INPROCEEDINGS{hino2024physical,
  author={Hino, Kanta and Tanaka, Shu},
  booktitle={2024 IEEE International Conference on Quantum Computing and Engineering (QCE)}, 
  title={Physical Properties of Error Reduction Algorithms for {I}sing Machines}, 
  year={2024},
  volume={02},
  number={},
  pages={434-435},
  doi={10.1109/QCE60285.2024.10342}
}

@article{hattori2025advantages,
  title={Advantages of fixing spins in quantum annealing},
  author={Hattori, Tomohiro and Irie, Hirotaka and Kadowaki, Tadashi and Tanaka, Shu},
  journal={J. Phys. Soc. Jpn.},
  volume={94},
  number={1},
  pages={013001},
  year={2025},
  publisher={The Physical Society of Japan}
}

@article{hattori2025impact,
  title={Impact of Fixing Spins in a Quantum Annealer with Energy Rescaling},
  author={Hattori, Tomohiro and Irie, Hirotaka and Kadowaki, Tadashi and Tanaka, Shu},
  journal={J. Phys. Soc. Jpn.},
  volume={94},
  number={7},
  pages={074001},
  year={2025},
  publisher={The Physical Society of Japan}
}

@article{hirama2023efficient,
  title={Efficient algorithm for binary quadratic problem by column generation and quantum annealing},
  author={Hirama, Sota and Ohzeki, Masayuki},
  journal={J. Phys. Soc. Jpn.},
  volume={92},
  number={11},
  pages={113002},
  year={2023},
  publisher={The Physical Society of Japan}
}

@ARTICLE{kanai2024annealing,
  author={Kanai, H. and Yamashita, M. and Tanahashi, K. and Tanaka, S.},
  journal={IEEE Access}, 
  title={Annealing-Assisted Column Generation for Inequality-Constrained Combinatorial Optimization Problems}, 
  year={2024},
  volume={12},
  number={},
  pages={157669-157685},
  doi={10.1109/ACCESS.2024.3486768}
}

@article{raymond2025quantum,
  title={Quantum error mitigation in quantum annealing},
  author={Raymond, Jack and Amin, Mohammad H and King, Andrew D and Harris, Richard and Bernoudy, William and Berkley, Andrew J and Boothby, Kelly and Smirnov, Anatoly and Altomare, Fabio and Babcock, Michael and others},
  journal={npj Quantum Information},
  volume={11},
  number={1},
  pages={38},
  year={2025},
  publisher={Nature Publishing Group UK London}
}

@article{kikuchi2023hybrid,
	author = {Kikuchi ,S. and Togawa ,N. and Tanaka ,S.},
	doi = {10.7566/JPSJ.92.124002},
	eprint = {https://doi.org/10.7566/JPSJ.92.124002},
	journal = {J. Phys. Soc. Jpn.},
	number = {12},
	pages = {124002},
	title = {Hybrid Optimization Method Using Simulated-Annealing-Based {I}sing Machine and Quantum Annealer},
	url = {https://doi.org/10.7566/JPSJ.92.124002},
	volume = {92},
	year = {2023}}

@article{shingu2024quantummitigation,
  title={Quantum annealing with error mitigation},
  author={Shingu, Yuta and Nikuni, Tetsuro and Kawabata, Shiro and Matsuzaki, Yuichiro},
  journal={Phys. Rev. A},
  volume={109},
  number={4},
  pages={042606},
  year={2024},
  publisher={APS}
}

@article{Irie2021Hybrid,
author = {Irie, H. and Liang, H. and Doi, T. and Gongyo, S. and Hatsuda, T.},
date = {2021/04/19},
doi = {10.1038/s41598-021-87676-z},
id = {Irie2021},
isbn = {2045-2322},
journal = {Sci. Rep.},
number = {1},
pages = {8426},
title = {Hybrid quantum annealing via molecular dynamics},
url = {https://doi.org/10.1038/s41598-021-87676-z},
volume = {11},
year = {2021}}

@article{Karmi2017,
author = {Karimi, H. and Rosenberg, G.},
date = {2017/05/16},
doi = {10.1007/s11128-017-1615-x},
id = {Karimi2017},
isbn = {1573-1332},
journal = {Quantum Inf. Process.},
number = {7},
pages = {166},
title = {Boosting quantum annealer performance via sample persistence},
url = {https://doi.org/10.1007/s11128-017-1615-x},
volume = {16},
year = {2017}}

@article{Karimi2017_2,
  title = {Effective optimization using sample persistence: {A} case study on quantum annealers and various {M}onte {C}arlo optimization methods},
  author = {Karimi, H. and Rosenberg, G. and Katzgraber, H. G.},
  journal = {Phys. Rev. E},
  volume = {96},
  issue = {4},
  pages = {043312},
  numpages = {14},
  year = {2017},
  publisher = {American Physical Society},
  doi = {10.1103/PhysRevE.96.043312},
  url = {https://link.aps.org/doi/10.1103/PhysRevE.96.043312}
}

@ARTICLE{Atobe_2022,
author={Atobe, Y. and Tawada, M. and Togawa, N.},
journal={IEEE Trans. Comput.},
title={Hybrid Annealing Method Based on sub{Q}{U}{B}{O} Model Extraction With Multiple Solution Instances},
year={2022},
volume={71},
number={10},
pages={2606},
doi={10.1109/TC.2021.3138629}}

@article{pudenz2014error,
  title={Error-corrected quantum annealing with hundreds of qubits},
  author={Pudenz, Kristen L and Albash, Tameem and Lidar, Daniel A},
  journal={Nat. Commun.},
  volume={5},
  number={1},
  pages={3243},
  year={2014},
  publisher={Nature Publishing Group UK London}
}

@article{pudenz2015quantumerror,
  title={Quantum annealing correction for random {I}sing problems},
  author={Pudenz, Kristen L and Albash, Tameem and Lidar, Daniel A},
  journal={Phys. Rev. A},
  volume={91},
  number={4},
  pages={042302},
  year={2015},
  publisher={APS}
}

@article{matsuura2016mean,
  title={Mean field analysis of quantum annealing correction},
  author={Matsuura, Shunji and Nishimori, Hidetoshi and Albash, Tameem and Lidar, Daniel A},
  journal={Phys. Rev. Lett.},
  volume={116},
  number={22},
  pages={220501},
  year={2016},
  publisher={APS}
}

@article{mishra2016performance,
  title={Performance of two different quantum annealing correction codes},
  author={Mishra, Anurag and Albash, Tameem and Lidar, Daniel A},
  journal={Quantum Inf. Process.},
  volume={15},
  number={2},
  pages={609--636},
  year={2016},
  publisher={Springer}
}

@article{vinci2015quantumerror,
  title={Quantum annealing correction with minor embedding},
  author={Vinci, Walter and Albash, Tameem and Paz-Silva, Gerardo and Hen, Itay and Lidar, Daniel A},
  journal={Phys. Rev. A},
  volume={92},
  number={4},
  pages={042310},
  year={2015},
  publisher={APS}
}

@article{vinci2016nested,
  title={Nested quantum annealing correction},
  author={Vinci, Walter and Albash, Tameem and Lidar, Daniel A},
  journal={npj Quantum Inf.},
  volume={2},
  number={1},
  pages={1--6},
  year={2016},
  publisher={Nature Publishing Group}
}

@article{matsuura2017quantum,
  title={Quantum-annealing correction at finite temperature: {F}erromagnetic p-spin models},
  author={Matsuura, Shunji and Nishimori, Hidetoshi and Vinci, Walter and Albash, Tameem and Lidar, Daniel A},
  journal={Phys. Rev. A},
  volume={95},
  number={2},
  pages={022308},
  year={2017},
  publisher={APS}
}

@article{vinci2018scalable,
  title={Scalable effective-temperature reduction for quantum annealers via nested quantum annealing correction},
  author={Vinci, Walter and Lidar, Daniel A},
  journal={Phys. Rev. A},
  volume={97},
  number={2},
  pages={022308},
  year={2018},
  publisher={APS}
}

@article{matsuura2019nested,
  title={Nested quantum annealing correction at finite temperature: p-spin models},
  author={Matsuura, Shunji and Nishimori, Hidetoshi and Vinci, Walter and Lidar, Daniel A},
  journal={Phys. Rev. A},
  volume={99},
  number={6},
  pages={062307},
  year={2019},
  publisher={APS}
}

@article{pearson2019analog,
  title={Analog errors in quantum annealing: doom and hope},
  author={Pearson, Adam and Mishra, Anurag and Hen, Itay and Lidar, Daniel A},
  journal={npj Quantum Inf.},
  volume={5},
  number={1},
  pages={107},
  year={2019},
  publisher={Nature Publishing Group UK London}
}

@article{li2019improved,
  title={Improved {B}oltzmann machines with error corrected quantum annealing},
  author={Li, Richard Y and Albash, Tameem and Lidar, Daniel A},
  journal={arXiv preprint arXiv:1910.01283},
  year={2019}
}

@article{li2020limitations,
  title={Limitations of error corrected quantum annealing in improving the performance of {B}oltzmann machines},
  author={Li, Richard Y and Albash, Tameem and Lidar, Daniel A},
  journal={Quantum Sci. Technol.},
  volume={5},
  number={4},
  pages={045010},
  year={2020},
  publisher={IOP Publishing}
}

@article{bennett2023using,
  title={Using copies can improve precision in continuous-time quantum computing},
  author={Bennett, Jemma and Callison, Adam and O’Leary, Tom and West, Mia and Chancellor, Nicholas and Kendon, Viv},
  journal={Quantum Sci. Technol.},
  volume={8},
  number={3},
  pages={035031},
  year={2023},
  publisher={IOP Publishing}
}

@article{Pelofske_2022,
author = {Pelofske, E. and Hahn, G. and Djidjev, H. N.},
date = {2022/03/16},
doi = {10.1038/s41598-022-08394-8},
id = {Pelofske2022},
isbn = {2045-2322},
journal = {Sci. Rep.},
number = {1},
pages = {4499},
title = {Parallel quantum annealing},
url = {https://doi.org/10.1038/s41598-022-08394-8},
volume = {12},
year = {2022}}

@article{altshuler2010anderson,
  title={Anderson localization makes adiabatic quantum optimization fail},
  author={Altshuler, Boris and Krovi, Hari and Roland, J{\'e}r{\'e}mie},
  journal={Proceedings of the National Academy of Sciences},
  volume={107},
  number={28},
  pages={12446--12450},
  year={2010},
  publisher={National Acad Sciences}
}

@article{roberts2020noise,
  title={Noise amplification at spin-glass bottlenecks of quantum annealing: {A} solvable model},
  author={Roberts, David and Cincio, Lukasz and Saxena, Avadh and Petukhov, Andre and Knysh, Sergey},
  journal={Phys. Rev. A},
  volume={101},
  number={4},
  pages={042317},
  year={2020},
  publisher={APS}
}

@article{Zaborniak:2021aql,
    author = "Zaborniak, T. and Sousa, R.",
    title = "{Benchmarking {H}amiltonian Noise in the {D}-{W}ave Quantum Annealer}",
    doi = "10.1109/TQE.2021.3050449",
    journal = "IEEE Trans. Quantum Eng.",
    volume = "2",
    pages = "3100206",
    year = "2021"
}

@article{somma2012quantum,
  title={Quantum speedup by quantum annealing},
  author={Somma, R. D and Nagaj, D. and Kieferov{\'a}, M.},
  journal={Phys. Rev. Lett.},
  volume={109},
  number={5},
  pages={050501},
  year={2012},
  publisher={APS}
}

@article{seki2012quantum,
  title={Quantum annealing with antiferromagnetic fluctuations},
  author={Seki, Y. and Nishimori, H.},
  journal={Phys. Rev. E},
  volume={85},
  number={5},
  pages={051112},
  year={2012},
  publisher={APS}
}

@article{susa2018quantum,
  title={Quantum annealing of the p-spin model under inhomogeneous transverse field driving},
  author={Susa, Y. and Yamashiro, Y. and Yamamoto, M. and Hen, I. and Lidar, D. A. and Nishimori, H.},
  journal={Phys. Rev. A},
  volume={98},
  number={4},
  pages={042326},
  year={2018},
  publisher={APS}
}

@article{susa2018exponential,
  title={Exponential speedup of quantum annealing by inhomogeneous driving of the transverse field},
  author={Susa, Y. and Yamashiro, Y. and Yamamoto, M. and Nishimori, H.},
  journal={J. Phys. Soc. Jpn.},
  volume={87},
  number={2},
  pages={023002},
  year={2018},
  publisher={The Physical Society of Japan}
}

@article{adame2020inhomogeneous,
  title={Inhomogeneous driving in quantum annealers can result in orders-of-magnitude improvements in performance},
  author={Adame, J. I. and McMahon, P. L.},
  journal={Quantum Sci. Technol.},
  volume={5},
  number={3},
  pages={035011},
  year={2020},
  publisher={IOP Publishing}
}

@article{albash2021diagonal,
  title={Diagonal catalysts in quantum adiabatic optimization},
  author={Albash, T. and Kowalsky, Ma.},
  journal={Phys. Rev. A},
  volume={103},
  number={2},
  pages={022608},
  year={2021},
  publisher={APS}
}

@article{feinstein2024effects,
  title={Effects of {XX} catalysts on quantum annealing spectra with perturbative crossings},
  author={Feinstein, N. and Fry-Bouriaux, L. and Bose, S. and Warburton, P. A.},
  journal={Phys. Rev. A},
  volume={110},
  number={4},
  pages={042609},
  year={2024},
  publisher={APS}
}

@article{ghosh2026enhancement,
  title={Enhancement of quantum annealing via n-local catalysts},
  author={Ghosh, Roopayan and Nutricati, Luca A and Feinstein, Natasha and Warburton, PA and Bose, Sougato},
  journal={Physical Review Research},
  volume={8},
  number={1},
  pages={013301},
  year={2026},
  publisher={APS}
}

@article{wang2025exponential,
  title={From exponential to quadratic: optimal control for a frustrated Ising ring model},
  author={Wang, Ruiyi and Roberto Arezzo, Vincenzo and Thengil, Kiran and Pecci, Giovanni and Santoro, Giuseppe E},
  journal={Quantum Science and Technology},
  volume={10},
  number={3},
  pages={035052},
  year={2025},
  publisher={IOP Publishing}
}

@article{kirkpatrick1983optimization,
  title={Optimization by simulated annealing},
  author={Kirkpatrick, Scott and Gelatt Jr, C Daniel and Vecchi, Mario P},
  journal={Science},
  volume={220},
  number={4598},
  pages={671--680},
  year={1983},
  publisher={American association for the advancement of science}
}

@article{albash2015decoherence,
  title={Decoherence in adiabatic quantum computation},
  author={Albash, Tameem and Lidar, Daniel A},
  journal={Phys. Rev. A},
  volume={91},
  number={6},
  pages={062320},
  year={2015},
  publisher={APS}
}

@article{young2013adiabatic,
  title={Adiabatic quantum optimization with the wrong Hamiltonian},
  author={Young, Kevin C and Blume-Kohout, Robin and Lidar, Daniel A},
  journal={Phys. Rev. A},
  volume={88},
  number={6},
  pages={062314},
  year={2013},
  publisher={APS}
}

@misc{D-Wave_params,
    author = {},
    title = {{P}er-{QPU} Solver Properties and Schedules},
    publisher = {{D}-{W}ave {S}ystems},
    howpublished = {\url{https://docs.dwavequantum.com/en/latest/quantum_research/solver_properties_specific.html#advantage2-system1-10}},
    note = {(accessed 2025-1-20)}
}

@article{Choi_2008,
author = {Choi, V.},
date = {2008/10/01},
doi = {10.1007/s11128-008-0082-9},
id = {Choi2008},
isbn = {1573-1332},
journal = {Quantum Inf. Process.},
number = {5},
pages = {193},
title = {Minor-embedding in adiabatic quantum computation: {I}. {T}he parameter setting problem},
url = {https://doi.org/10.1007/s11128-008-0082-9},
volume = {7},
year = {2008}}

@article{Choi_2011,
author = {Choi, V.},
date = {2011/06/01},
doi = {10.1007/s11128-010-0200-3},
id = {Choi2011},
isbn = {1573-1332},
journal = {Quantum Inf. Process.},
number = {3},
pages = {343},
title = {Minor-embedding in adiabatic quantum computation: {II}. {M}inor-universal graph design},
url = {https://doi.org/10.1007/s11128-010-0200-3},
volume = {10},
year = {2011}}

@misc{D-Wave_zephr,
    author ={Boothby, K. and King,A.D. and Raymond, J.},
    title={Zephyr Topology of {D}-{W}ave Quantum Processors},
    year ={2021},
    publisher={D-{W}ave {T}echnical report},
    url={https://www.dwavesys.com/media/2uznec4s/14-1056a-a\_zephyr\_topology\_of\_d-wave\_quantum\_processors.pdf},
}

@article{tamura2026black,
  title={Black-box optimization using factorization and Ising machines},
  author={Tamura, Ryo and Seki, Yuya and Minamoto, Yuki and Kitai, Koki and Matsuda, Yoshiki and Tanaka, Shu and Tsuda, Koji},
  journal={Applied Physics Reviews},
  volume={13},
  number={2},
  year={2026},
  publisher={AIP Publishing}
}

@article{hattori2026improving,
  title={Improving the efficiency of quantum annealing with controlled diagonal catalysts},
  author={Hattori, Tomohiro and Tanaka, Shu},
  journal={Physical Review A},
  volume={113},
  number={2},
  pages={022433},
  year={2026},
  publisher={APS}
}

@article{couzinie2025machine,
  title={Machine learning supported annealing for prediction of grand canonical crystal structures},
  author={Couzini{\'e}, Yannick and Seki, Yuya and Nishiya, Yusuke and Nishi, Hirofumi and Kosugi, Taichi and Tanaka, Shu and Matsushita, Yu-ichiro},
  journal={J. Phys. Soc. Jpn.},
  volume={94},
  number={4},
  pages={044802},
  year={2025},
  publisher={The Physical Society of Japan}
}

@article{harris2018phase,
  title={Phase transitions in a programmable quantum spin glass simulator},
  author={Harris, Richard and Sato, Yuki and Berkley, Andrew J and Reis, M and Altomare, Fabio and Amin, MH and Boothby, Kelly and Bunyk, P and Deng, C and Enderud, Colin and others},
  journal={Science},
  volume={361},
  number={6398},
  pages={162--165},
  year={2018},
  publisher={American Association for the Advancement of Science}
}

@article{king2018observation,
  title={Observation of topological phenomena in a programmable lattice of 1,800 qubits},
  author={King, Andrew D and Carrasquilla, Juan and Raymond, Jack and Ozfidan, Isil and Andriyash, Evgeny and Berkley, Andrew and Reis, Mauricio and Lanting, Trevor and Harris, Richard and Altomare, Fabio and others},
  journal={Nature},
  volume={560},
  number={7719},
  pages={456--460},
  year={2018},
  publisher={Nature Publishing Group UK London}
}

@article{honda2024development,
  title={Development of optimization method for truss structure by quantum annealing},
  author={Honda, Rio and Endo, Katsuhiro and Kaji, Taichi and Suzuki, Yudai and Matsuda, Yoshiki and Tanaka, Shu and Muramatsu, Mayu},
  journal={Sci. Rep.},
  volume={14},
  number={1},
  pages={13872},
  year={2024},
  publisher={Nature Publishing Group UK London}
}

@article{endo2022phase,
  title={A phase-field model by an {I}sing machine and its application to the phase-separation structure of a diblock polymer},
  author={Endo, Katsuhiro and Matsuda, Yoshiki and Tanaka, Shu and Muramatsu, Mayu},
  journal={Sci. Rep.},
  volume={12},
  number={1},
  pages={10794},
  year={2022},
  publisher={Nature Publishing Group UK London}
}

@article{xu2025quantum,
  title={Quantum annealing-assisted lattice optimization},
  author={Xu, Zhihao and Shang, Wenjie and Kim, Seongmin and Lee, Eungkyu and Luo, Tengfei},
  journal={npj Comput. Mater.},
  volume={11},
  number={1},
  pages={4},
  year={2025},
  publisher={Nature Publishing Group UK London}
}

@article{redfield1957theory,
  title={On the theory of relaxation processes},
  author={Redfield, Alfred G},
  journal={IBM J. Res. Dev.},
  volume={1},
  number={1},
  pages={19--31},
  year={1957},
  publisher={IBM}
}

@article{santoro2002theory,
  title={Theory of quantum annealing of an {I}sing spin glass},
  author={Santoro, Giuseppe E and Marton{\'a}k, Roman and Tosatti, Erio and Car, Roberto},
  journal={Science},
  volume={295},
  number={5564},
  pages={2427--2430},
  year={2002},
  publisher={American Association for the Advancement of Science}
}

@article{King_2022,
   title={Coherent quantum annealing in a programmable 2,000 qubit Ising chain},
   volume={18},
   ISSN={1745-2481},
   url={http://dx.doi.org/10.1038/s41567-022-01741-6},
   DOI={10.1038/s41567-022-01741-6},
   number={11},
   journal={Nat. Phys.},
   publisher={Springer Science and Business Media LLC},
   author={King, Andrew D. and Suzuki, Sei and Raymond, Jack and Zucca, Alex and Lanting, Trevor and Altomare, Fabio and Berkley, Andrew J. and Ejtemaee, Sara and Hoskinson, Emile and Huang, Shuiyuan and Ladizinsky, Eric and MacDonald, Allison J. R. and Marsden, Gaelen and Oh, Travis and Poulin-Lamarre, Gabriel and Reis, Mauricio and Rich, Chris and Sato, Yuki and Whittaker, Jed D. and Yao, Jason and Harris, Richard and Lidar, Daniel A. and Nishimori, Hidetoshi and Amin, Mohammad H.},
   year={2022},
   pages={1324–1328} }
\end{document}